%% file: main.tex
\documentclass[conference]{IEEEtran}
\IEEEoverridecommandlockouts

\usepackage[nocompress]{cite}   
\usepackage{amsmath,amssymb,amsfonts}
\usepackage{graphicx}
\usepackage{booktabs}
\usepackage{algorithm}
\usepackage{algpseudocode}
\usepackage{multirow}
\usepackage{array}
\usepackage{tabularx}
\usepackage{threeparttable}
\usepackage{pifont}
\usepackage{microtype}
\usepackage[table]{xcolor}
\usepackage{hyperref}
    
\definecolor{TabRule}{HTML}{52514E}
\definecolor{TabStoneHead}{HTML}{E6E5E1}
\definecolor{TabStoneRow}{HTML}{F6F5F2}
\definecolor{TabBlueHead}{HTML}{DCE8F6}
\definecolor{TabBlueCol}{HTML}{EAF2FB}
\definecolor{TabBlueRow}{HTML}{F4F8FD}
\definecolor{AlgBlue}{HTML}{2A78D6}

\newcommand{\algbox}[1]{%
  {\setlength{\fboxsep}{1.5pt}\colorbox{TabBlueCol}{\strut #1}}%
}
\algrenewcommand{\algorithmiccomment}[1]{%
  \hfill\textcolor{TabRule}{\textit{// #1}}%
}

\definecolor{cHead}{HTML}{EEF3FA}
\definecolor{cHeadText}{HTML}{22304F}
\definecolor{cBase}{HTML}{5B6B82}
\definecolor{cLearn}{HTML}{7A4FD6}
\definecolor{cMap}{HTML}{1F7AE0}
\definecolor{cPub}{HTML}{E8562E}
\definecolor{cRef}{HTML}{149C8A}
\definecolor{cYes}{HTML}{18A558}
\definecolor{cOpt}{HTML}{F29A0C}
\definecolor{cOn}{HTML}{0EA5E9}
\definecolor{cOff}{HTML}{DB2777}

\providecommand{\Y}{\textcolor{cYes}{\ding{51}}}
\providecommand{\Opt}{\textcolor{cOpt}{\ding{109}}}
\providecommand{\N}{\textcolor{black!25}{--}}

\providecommand{\cdotmark}[1]{%
  \textcolor{#1}{\raisebox{0.15ex}{\tiny\ding{108}}}%
  \hspace{2.5pt}%
}
\providecommand{\On}{\cdotmark{cOn}online}
\providecommand{\Off}{\cdotmark{cOff}offline}
\providecommand{\Del}{\cdotmark{cOpt}delayed $\le$3\,s}

\providecommand{\hd}[1]{\textcolor{cHeadText}{\textbf{#1}}}
\providecommand{\hdrule}{%
  \leaders\hrule height 2.6pt depth -2.2pt
  \hskip 0pt plus 0.5fill\relax
}
\providecommand{\hdspan}[1]{%
  \hd{\hdrule\enspace #1\enspace\hdrule}%
  \hskip 0pt plus -1fill\kern0pt
}

\newcolumntype{L}[1]{%
  >{\hsize=#1\hsize\raggedright\arraybackslash}X%
}

\providecommand{\grp}[2]{%
  \rowcolor{#1!18}%
  \multicolumn{10}{l}{%
    \textbf{\textcolor{#1!78!black}{#2}}%
  }\\
}
\providecommand{\tint}[1]{\rowcolor{#1!4}}

\hypersetup{colorlinks=true, urlcolor=blue, linkcolor=black, citecolor=black}

\input{macros_numbers.tex}   

\begin{document}

\title{CANTraj: Recovering Vehicle Trajectories from CAN Signals in Naturalistic Driving}

\author{\IEEEauthorblockN{Yuhang Wang\textsuperscript{1}, Jingxin Yang\textsuperscript{2}, Chuheng Wei\textsuperscript{3}, Suyang Chen\textsuperscript{4}, Haozhuang Chi\textsuperscript{5} and Hao Zhou\textsuperscript{1,}\IEEEauthorrefmark{1}\thanks{\IEEEauthorrefmark{1}Corresponding author: Hao Zhou (haozhou1@usf.edu).}}
\IEEEauthorblockA{\textsuperscript{1}University of South Florida, USA \quad \textsuperscript{2}Stanford University, USA \quad \textsuperscript{3}Purdue University, USA \\ \textsuperscript{4}University of Florida, USA \quad \textsuperscript{5}Nanyang Technological University, Singapore}
\IEEEauthorblockA{yuhangw@usf.edu, jingxiny@stanford.edu, wch@purdue.edu, \\ suyang.chen@ufl.edu, chi0001@e.ntu.edu.sg, haozhou1@usf.edu}}

\maketitle

\begin{abstract}
\input{sections/abstract.tex}
\end{abstract}

\input{sections/introduction.tex}

\input{sections/related.tex}
\input{sections/problem.tex}
\input{sections/data.tex}
\input{sections/methods.tex}

\input{sections/results.tex}

\input{sections/discussion.tex}

\clearpage

\paragraph*{Generative AI use}

Claude Code was used to organize the code for supplementary. ChatGPT (OpenAI Codex) assisted with the final LaTeX typesetting. Neither system generated or modified the experimental results and paper content. Additionally, the open-weight model Qwen3-14B (Alibaba, provided via vLLM) was the system evaluated in the proxy study in Section VI; this model did not generate any text, code, or analysis for this paper.


\bibliographystyle{IEEEtran}
\bibliography{refs}
\end{document}

%% file: macros_numbers.tex
\newcommand{\AThreeDrives}{60}
\newcommand{\AThreeIdenticalPct}{100}

\newcommand{\AThreeSeedMovedPctFive}{19}
\newcommand{\AThreeWindows}{651}

\newcommand{\AgentBeamTestRoute}{69}

\newcommand{\AgentCohortAllN}{659}

\newcommand{\AgentCohortFewShotN}{391}

\newcommand{\AgentCohortTestN}{268}

\newcommand{\AgentLlmTestCost}{36.4}

\newcommand{\AgentLlmTestRoute}{75}

\newcommand{\AgentMaxLlmFewShotRoute}{81}

\newcommand{\AgentMaxLlmTestRoute}{78}

\newcommand{\AgentMaxOracleTestRoute}{87}

\newcommand{\AgentMaxPairLlmVsPrevAllRouteDiff}{$-$0.4}

\newcommand{\AgentMaxPairLlmVsPrevAllRouteHi}{2.0}
\newcommand{\AgentMaxPairLlmVsPrevAllRouteLo}{$-$2.8}

\newcommand{\AgentMaxPairLlmVsSelTestRouteDiff}{4.1}

\newcommand{\AgentMaxPairLlmVsSelTestRouteHi}{7.6}
\newcommand{\AgentMaxPairLlmVsSelTestRouteLo}{1.3}

\newcommand{\AgentMaxVoteFewShotRoute}{83}

\newcommand{\AgentMaxVoteTestRoute}{79}

\newcommand{\AgentOracleTestRoute}{83}

\newcommand{\AgentPairLlmVsSelFewShotRouteDiff}{0.7}

\newcommand{\AgentPairLlmVsSelFewShotRouteHi}{2.8}
\newcommand{\AgentPairLlmVsSelFewShotRouteLo}{$-$2.6}

\newcommand{\AgentPairLlmVsSelTestRouteDiff}{0.1}

\newcommand{\AgentPairLlmVsSelTestRouteHi}{3.4}
\newcommand{\AgentPairLlmVsSelTestRouteLo}{$-$1.6}

\newcommand{\AgentSelTestCost}{1.1}

\newcommand{\AgentSelTestRoute}{75}

\newcommand{\AreaCtTestOutPct}{74}

\newcommand{\AreaSpyTestOutPct}{80}

\newcommand{\CFailDensHighPct}{2.9}

\newcommand{\CFailDensLowPct}{0.9}

\newcommand{\CFailTurnsSevenPlusN}{402}

\newcommand{\CFailTurnsSevenPlusPct}{8.7}

\newcommand{\CFailTurnsZeroOnePct}{0.1}

\newcommand{\CohElasticPathingFMHmFifteenTestFail}{11}
\newcommand{\CohElasticPathingFMHmFifteenTestMed}{3{,}335}

\newcommand{\CohMacrotrackingOMHpFifteenTestMed}{1{,}351}

\newcommand{\CohMapBeamNoneFMHmFifteenTestMed}{5{,}352}

\newcommand{\CohMapBeamSteerFMHpFifteenTestMed}{303}

\newcommand{\CohMapBeamSteerFMHpFiveTestMed}{44}

\newcommand{\CohMapBeamSteerFMHpOneTestMed}{17}

\newcommand{\CohMapPfSteerOMHpFifteenTestMed}{672}

\newcommand{\CohMapPfSteerOMHpFiveTestMed}{56}

\newcommand{\CohMapPfSteerOMHpOneTestMed}{19}

\newcommand{\CohOpdTwoFMHpFifteenTestMed}{1{,}316}

\newcommand{\CohOpdTwoOnlineOMHpOneTestMed}{11}

\newcommand{\CohPairBeamVsLineageFifteenTestMean}{$-$547}
\newcommand{\CohPairBeamVsLineageFifteenTestMeanHi}{$-$109}
\newcommand{\CohPairBeamVsLineageFifteenTestMeanLo}{$-$1{,}390}

\newcommand{\CohPairBeamVsPfFifteenTestHi}{$-$1.6}
\newcommand{\CohPairBeamVsPfFifteenTestLo}{$-$6.1}
\newcommand{\CohPairBeamVsPfFifteenTestMean}{$-$552}
\newcommand{\CohPairBeamVsPfFifteenTestMeanHi}{$-$98.6}
\newcommand{\CohPairBeamVsPfFifteenTestMeanLo}{$-$1{,}421}
\newcommand{\CohPairBeamVsPfFifteenTestMed}{$-$3.1}

\newcommand{\CohTestDrivers}{37}
\newcommand{\CohTestN}{205}

\newcommand{\CompDFewShotOracleRoutePct}{98}

\newcommand{\CompDTestAdeGainHi}{$-$2.3}
\newcommand{\CompDTestAdeGainLo}{$-$5.4}

\newcommand{\CompDTestAdeGainMed}{$-$3.4}
\newcommand{\CompDTestAllFailHi}{3.6}
\newcommand{\CompDTestAllFailLo}{1.4}
\newcommand{\CompDTestAllFailPct}{2.3}

\newcommand{\CompDTestOracleRoutePct}{98}

\newcommand{\CompDTestRefRoutePct}{88}
\newcommand{\CompDTestRouteGainHi}{16}
\newcommand{\CompDTestRouteGainLo}{6}
\newcommand{\CompDTestRouteGainPct}{10}

\newcommand{\CompMemDTestBeamAttainPct}{19}

\newcommand{\CompMemDTestLineageAttainPct}{22}

\newcommand{\CompMemDTestOpdTwoAttainPct}{22}

\newcommand{\ConfBeamHpFiveFewShotAuroc}{0.49}

\newcommand{\ConfBeamHpFiveTestAuroc}{0.47}
\newcommand{\ConfBeamHpFiveTestAurocHi}{0.52}
\newcommand{\ConfBeamHpFiveTestAurocLo}{0.41}

\newcommand{\ConfBeamHpFiveTestNinetyCovPct}{46}
\newcommand{\ConfBeamHpFiveTestNinetyErrPct}{13}

\newcommand{\ConfBeamHpFiveTestNinetyRoutePct}{87}

\newcommand{\ConfBeamHpFiveTestProbOnePct}{29}

\newcommand{\ConfBeamHpFiveTestRho}{$-$0.01}

\newcommand{\ConfBeamNoneHmFiveTestAuroc}{0.65}

\newcommand{\ConfBeamNoneHpFiveTestAuroc}{0.81}

\newcommand{\ConfBeamNoneHpFiveTestNinetyCovPct}{13}

\newcommand{\ConfBeamNoneHpFiveTestNinetyRoutePct}{97}

\newcommand{\ConfProbOneDewriHpPct}{87}

\newcommand{\ConfProbOneEpHpPct}{69}

\newcommand{\CostDrSteerFive}{0.03}

\newcommand{\CostElasticPathingFiveNinety}{0.13}

\newcommand{\CostGraph}{2.6}

\newcommand{\CostMacrotrackingFive}{0.08}

\newcommand{\CostMapBeamSteerFive}{0.24}

\newcommand{\CostMapPfSteerFive}{0.57}

\newcommand{\CostN}{92}
\newcommand{\CostNFive}{76}

\newcommand{\CostOpdTwoFive}{0.08}

\newcommand{\CostPedalCanTraceAreaFive}{0.09}

\newcommand{\CostPedalCanTraceFive}{0.63}

\newcommand{\CostPedalNFive}{25}
\newcommand{\CostRssMax}{2.1}
\newcommand{\CostSpyAreaFailFive}{9}

\newcommand{\CostSpyAreaFive}{4.93}

\newcommand{\CtDiagIntersectionsMed}{14}

\newcommand{\CtDiagTraceNodesMed}{6}
\newcommand{\DensCutHigh}{246}
\newcommand{\DensCutLow}{56}

\newcommand{\DensMapBeamSteerHighMedFifteen}{162}

\newcommand{\DensMapBeamSteerLowMedFifteen}{77}

\newcommand{\DensMapPfNoneHighMed}{843}

\newcommand{\DensMapPfNoneLowMed}{73}

\newcommand{\DensMapPfNoneMidMed}{677}

\newcommand{\DensMapPfSteerHighDrivers}{81}
\newcommand{\DensMapPfSteerHighLostPct}{27}
\newcommand{\DensMapPfSteerHighMed}{36}
\newcommand{\DensMapPfSteerHighMedFifteen}{264}
\newcommand{\DensMapPfSteerLowDrivers}{33}
\newcommand{\DensMapPfSteerLowLostPct}{17}
\newcommand{\DensMapPfSteerLowMed}{29}
\newcommand{\DensMapPfSteerLowMedFifteen}{89}
\newcommand{\DensMapPfSteerMidDrivers}{82}

\newcommand{\DensMapPfSteerMidMed}{39}

\newcommand{\DensNFifteen}{374}
\newcommand{\DensNFive}{1{,}292}

\newcommand{\DiagLowTurnPrecSpyHp}{1.00}

\newcommand{\DiagLowTurnRecallSpyHp}{0.36}

\newcommand{\DrvRouteBeamTestBelowHalfPct}{3}
\newcommand{\DrvRouteBeamTestMed}{88}

\newcommand{\DrvRouteBeamTestNDriversGeFive}{36}

\newcommand{\FailBeamAfterTurnPct}{38}
\newcommand{\FailBeamChancePct}{22}
\newcommand{\FailBeamLostPct}{70}

\newcommand{\FailPfAfterTurnPct}{43}
\newcommand{\FailPfChancePct}{23}
\newcommand{\FailPfLostPct}{73}
\newcommand{\FailPfN}{1{,}169}
\newcommand{\FailPfNoneAfterTurnPct}{23}
\newcommand{\FailPfNoneChancePct}{22}
\newcommand{\FailPfNoneLostPct}{93}

\newcommand{\FullBeyondDiscFewShot}{36}
\newcommand{\FullBeyondDiscTest}{18}
\newcommand{\FullDurMedFewShot}{11}
\newcommand{\FullDurMedTest}{19}

\newcommand{\FullMapOkFewShot}{454}
\newcommand{\FullMapOkTest}{282}

\newcommand{\MacroSampleMedian}{7.1}
\newcommand{\MapStartDistMedian}{2.6}
\newcommand{\MapStartMissingMain}{31}
\newcommand{\NBrands}{27}
\newcommand{\NCohortDrivers}{81}
\newcommand{\NCohortDrives}{426}
\newcommand{\NDrivers}{431}
\newcommand{\NDrives}{3{,}675}
\newcommand{\NDrivesQcom}{2{,}793}
\newcommand{\NDrivesUblox}{882}

\newcommand{\NEligibleInFolds}{97{,}340}
\newcommand{\NFewShotDrivers}{51}
\newcommand{\NFewShotDrives}{474}
\newcommand{\NFewShotFifteenMin}{565}
\newcommand{\NFewShotFiveMin}{1{,}963}
\newcommand{\NFewShotFullMin}{460}

\newcommand{\NHeldoutDrivers}{94}
\newcommand{\NHeldoutDrives}{776}

\newcommand{\NHours}{1{,}380}
\newcommand{\NKm}{85{,}919}
\newcommand{\NPedalFewShotDrives}{64}

\newcommand{\NPedalTestDrives}{48}
\newcommand{\NPlatformUnconfirmedDrivers}{32}
\newcommand{\NPlatformUnconfirmedDrives}{65}
\newcommand{\NPlatforms}{266}
\newcommand{\NRegionCA}{5}
\newcommand{\NRegionOther}{35}
\newcommand{\NRegionUS}{44}
\newcommand{\NReleaseDrivers}{454}
\newcommand{\NReleaseDrives}{3{,}940}
\newcommand{\NReleaseHours}{1{,}659}
\newcommand{\NReleaseNoAnchorDrives}{131}
\newcommand{\NReleaseNoWindowDrives}{69}
\newcommand{\NReleasePlatformDrives}{65}
\newcommand{\NReleaseUnusedDrives}{265}
\newcommand{\NTestDrivers}{43}
\newcommand{\NTestDrives}{302}
\newcommand{\NTestFifteenMin}{608}
\newcommand{\NTestFiveMin}{1{,}917}
\newcommand{\NTestFullMin}{282}

\newcommand{\NTilesTruncated}{149}
\newcommand{\NTilesVtwo}{5{,}111}

\newcommand{\NUnassignedDrivers}{86}
\newcommand{\NUnassignedDrives}{122}

\newcommand{\OpdSampleLogs}{8}
\newcommand{\OpdSampleMaxMedian}{0.33}

\newcommand{\PairBeamVsLineageFifteenFewShotMean}{$-$221}
\newcommand{\PairBeamVsLineageFifteenFewShotMeanHi}{$-$43.3}
\newcommand{\PairBeamVsLineageFifteenFewShotMeanLo}{$-$464}

\newcommand{\PairBeamVsLineageFiveTestHi}{0.3}
\newcommand{\PairBeamVsLineageFiveTestLo}{$-$1.3}

\newcommand{\PairBeamVsLineageFiveTestMed}{$-$0.5}

\newcommand{\PairBeamVsPfFiveTestHi}{$-$2.0}
\newcommand{\PairBeamVsPfFiveTestLo}{$-$3.0}

\newcommand{\PairBeamVsPfFiveTestMed}{$-$2.7}

\newcommand{\PairDewriVsEpFiveTestMean}{929}
\newcommand{\PairDewriVsEpFiveTestMeanHi}{1{,}417}
\newcommand{\PairDewriVsEpFiveTestMeanLo}{380}
\newcommand{\PairDewriVsEpFiveTestMed}{0.0}

\newcommand{\PairEpVsBeamNoneHmFiveTestHi}{$-$33.0}
\newcommand{\PairEpVsBeamNoneHmFiveTestLo}{$-$212}

\newcommand{\PairEpVsBeamNoneHmFiveTestMed}{$-$51.2}

\newcommand{\PairGruPfVsPfFiveFewShotMed}{5.4}

\newcommand{\PairGruPfVsPfFiveFewShotSeeds}{3}
\newcommand{\PairGruPfVsPfFiveFewShotSeedsSig}{2}

\newcommand{\PairGruPfVsPfFiveTestMed}{9.9}

\newcommand{\PairGruVsDrFiveFewShotMed}{$-$714}

\newcommand{\PairGruVsDrFiveFewShotSeedSd}{115}
\newcommand{\PairGruVsDrFiveFewShotSeeds}{3}
\newcommand{\PairGruVsDrFiveFewShotSeedsSig}{3}

\newcommand{\PairGruVsDrFiveTestHi}{361}
\newcommand{\PairGruVsDrFiveTestLo}{$-$592}

\newcommand{\PairGruVsDrFiveTestMed}{$-$120}

\newcommand{\PairGyroVsSteerFiveFewShotMed}{$-$38.9}
\newcommand{\PairGyroVsSteerFiveFewShotN}{732}

\newcommand{\PairGyroVsSteerFiveTestMed}{28.1}
\newcommand{\PairGyroVsSteerFiveTestN}{667}

\newcommand{\PairHeadBeamFiveTestMean}{$-$395}
\newcommand{\PairHeadBeamFiveTestMeanHi}{$-$206}
\newcommand{\PairHeadBeamFiveTestMeanLo}{$-$554}
\newcommand{\PairHeadBeamFiveTestMed}{0.0}

\newcommand{\PairHeadBeamFiveTestWinA}{16}

\newcommand{\PairHeadPfFiveTestHi}{$-$11.3}
\newcommand{\PairHeadPfFiveTestLo}{$-$41.4}
\newcommand{\PairHeadPfFiveTestMean}{$-$599}

\newcommand{\PairHeadPfFiveTestMed}{$-$27.5}

\newcommand{\PairLineageVsPfFiveTestHi}{$-$0.9}
\newcommand{\PairLineageVsPfFiveTestLo}{$-$2.3}

\newcommand{\PairLineageVsPfFiveTestMed}{$-$1.4}

\newcommand{\PairMacroCalVsMacroFiveTestMed}{$-$0.1}

\newcommand{\PairMacroFixFifteenTestMean}{$-$301}
\newcommand{\PairMacroFixFifteenTestMeanHi}{$-$147}
\newcommand{\PairMacroFixFifteenTestMeanLo}{$-$591}
\newcommand{\PairMacroFixFifteenTestMed}{0.0}

\newcommand{\PairMapGainFiveTestMed}{$-$825}

\newcommand{\PairMapGainFiveTestWinA}{89}

\newcommand{\PairOpdCalVsOpdFiveTestMed}{0.1}

\newcommand{\PairOpdFixFifteenTestMean}{$-$21.2}
\newcommand{\PairOpdFixFifteenTestMeanHi}{12.6}
\newcommand{\PairOpdFixFifteenTestMeanLo}{$-$54.2}

\newcommand{\PairOpdVsPfFiveTestHi}{2.3}
\newcommand{\PairOpdVsPfFiveTestLo}{$-$3.5}
\newcommand{\PairOpdVsPfFiveTestMean}{116}
\newcommand{\PairOpdVsPfFiveTestMeanHi}{245}
\newcommand{\PairOpdVsPfFiveTestMeanLo}{33.6}
\newcommand{\PairOpdVsPfFiveTestMed}{$-$1.9}

\newcommand{\PairResidPfVsPfFiveFewShotMed}{11.3}

\newcommand{\PairResidPfVsPfFiveFewShotSeedsSig}{3}

\newcommand{\PairResidPfVsPfFiveTestMed}{6.0}

\newcommand{\PairResidVsDrFiveFewShotMed}{$-$841}

\newcommand{\PairResidVsDrFiveFewShotSeedSd}{75.8}

\newcommand{\PairResidVsDrFiveFewShotSeedsSig}{3}

\newcommand{\PairSteerPfFiveTestMed}{$-$59.5}

\newcommand{\PairYawVsSteerFiveFewShotDrivers}{12}
\newcommand{\PairYawVsSteerFiveFewShotHi}{$-$66.8}
\newcommand{\PairYawVsSteerFiveFewShotLo}{$-$1{,}184}

\newcommand{\PairYawVsSteerFiveFewShotMed}{$-$338}
\newcommand{\PairYawVsSteerFiveFewShotN}{188}

\newcommand{\PairYawVsSteerFiveTestDrivers}{11}
\newcommand{\PairYawVsSteerFiveTestHi}{912}
\newcommand{\PairYawVsSteerFiveTestLo}{$-$2{,}305}

\newcommand{\PairYawVsSteerFiveTestMed}{519}
\newcommand{\PairYawVsSteerFiveTestN}{658}

\newcommand{\PrepTilesMin}{46}
\newcommand{\PrepValhallaGb}{3.4}
\newcommand{\PrepValhallaMin}{44}

\newcommand{\ResBeamFifteenTestMed}{149}

\newcommand{\ResBeamFiveTestDrv}{35}

\newcommand{\ResBeamFiveTestHi}{54}
\newcommand{\ResBeamFiveTestLo}{24}
\newcommand{\ResBeamFiveTestMed}{37}

\newcommand{\ResBeamFiveTestRoute}{88}

\newcommand{\ResBeamFullFewShotMed}{94}

\newcommand{\ResBeamFullTestMed}{1{,}179}

\newcommand{\ResBeamHmFifteenTestMed}{155}

\newcommand{\ResBeamHmFiveTestMed}{41}

\newcommand{\ResBeamHmFiveTestRoute}{81}

\newcommand{\ResBeamHmFullFewShotMed}{124}

\newcommand{\ResBeamHmFullTestMed}{1{,}244}

\newcommand{\ResBeamNoneFiveTestMed}{165}

\newcommand{\ResBeamNoneFiveTestRoute}{54}

\newcommand{\ResBeamNoneHmFiveTestMed}{1{,}314}

\newcommand{\ResCtAreaFiveTestOut}{85}

\newcommand{\ResCtFiveFewShotN}{162}
\newcommand{\ResCtFiveFewShotOut}{5}

\newcommand{\ResCtFiveTestN}{245}
\newcommand{\ResCtFiveTestOut}{2}

\newcommand{\ResCurvFiveTestMed}{84}

\newcommand{\ResDewriFiveTestMed}{293}

\newcommand{\ResDewriHpFiveTestMed}{81}

\newcommand{\ResDrCalFiveTestMed}{1{,}312}

\newcommand{\ResDrFiveFewShotDrivers}{50}

\newcommand{\ResDrFiveFewShotMed}{1{,}669}
\newcommand{\ResDrFiveFewShotN}{1{,}963}

\newcommand{\ResDrFiveTestMed}{921}

\newcommand{\ResEpFiveTestDrv}{538}

\newcommand{\ResEpFiveTestFail}{62}

\newcommand{\ResEpFiveTestMed}{129}
\newcommand{\ResEpFiveTestN}{1{,}917}

\newcommand{\ResEpHpFiveTestFail}{89}

\newcommand{\ResEpHpFiveTestMed}{87}

\newcommand{\ResGruFiveFewShotMed}{393}

\newcommand{\ResGruFiveTestMed}{543}

\newcommand{\ResMacroFiveFewShotMed}{52}

\newcommand{\ResMacroFiveTestMed}{84}

\newcommand{\ResOpdTwoFiveFewShotMed}{24}

\newcommand{\ResOpdTwoFiveTestMed}{42}

\newcommand{\ResOpdTwoFiveTestRoute}{79}

\newcommand{\ResOpdTwoOnFiveTestMed}{42}

\newcommand{\ResPfCurvFiveTestMed}{44}

\newcommand{\ResPfFifteenTestMed}{178}

\newcommand{\ResPfFiveFewShotN}{1{,}959}

\newcommand{\ResPfFiveTestDrivers}{42}
\newcommand{\ResPfFiveTestDrv}{41}

\newcommand{\ResPfFiveTestHi}{59}
\newcommand{\ResPfFiveTestLo}{29}
\newcommand{\ResPfFiveTestMed}{43}
\newcommand{\ResPfFiveTestN}{1{,}917}

\newcommand{\ResPfFiveTestRoute}{87}

\newcommand{\ResPfHmFiveTestMed}{138}

\newcommand{\ResPfHmFiveTestRoute}{79}

\newcommand{\ResPfNoneFiveTestMed}{286}

\newcommand{\ResPfNoneFiveTestRoute}{56}

\newcommand{\ResPfNoneHmFiveTestMed}{1{,}773}

\newcommand{\ResResidFiveFewShotMed}{435}

\newcommand{\ResSpyAreaFiveTestFail}{227}

\newcommand{\ResSpyFiveTestRoute}{11}

\newcommand{\ResSpyHpFiveTestRoute}{20}

\newcommand{\ResSpyHpFullFewShotRoute}{6}

\newcommand{\ResSpyHpFullTestRoute}{4}

\newcommand{\ResStraightFiveTestMed}{735}

\newcommand{\RobMapBeamNonePqFiveTestRouteDiff}{$-$0.6}
\newcommand{\RobMapBeamNonePqFiveTestRouteDiffHi}{0.1}
\newcommand{\RobMapBeamNonePqFiveTestRouteDiffLo}{$-$1.5}

\newcommand{\RobMapBeamPqFiveTestRouteDiff}{$-$3.3}

\newcommand{\RobMapBeamSFifthHzFewShotRouteDiff}{$-$1.8}

\newcommand{\RobMapHardMinRoute}{82.6}

\newcommand{\RobMapPfPqFiveFewShotRouteDiff}{$-$7.6}

\newcommand{\RobMapPfSFifthHzFewShotRouteDiff}{$-$5.6}

\newcommand{\RobMapPfSFifthHzTestMed}{49}
\newcommand{\RobMapPfSFifthHzTestMedClean}{43}

\newcommand{\RobMapPfSFifthHzTestRouteDiff}{$-$2.7}

\newcommand{\RobMapPfSqTwentyTestRouteDiff}{$-$1.4}
\newcommand{\RobMapPfSqTwentyTestRouteDiffHi}{$-$0.4}
\newcommand{\RobMapPfSqTwentyTestRouteDiffLo}{$-$2.5}

\newcommand{\RobMapPqOneMaxRouteAbs}{0.7}

\newcommand{\RobMapSteerMildMaxRouteAbs}{1.3}
\newcommand{\RobustDrMaxAbs}{0.4}

\newcommand{\SigGainMapFlagsSpeedHi}{10.7}
\newcommand{\SigGainMapFlagsSpeedLo}{0.8}

\newcommand{\SigGainMapFlagsSpeedMed}{5.2}

\newcommand{\SigGainMapPedAllMax}{0.9}

\newcommand{\SigGainMapPedAllMin}{$-$6.4}

\newcommand{\SigGainMapPedSpeedMax}{6.1}

\newcommand{\SigGainMapPedSpeedMin}{$-$4.8}

\newcommand{\SigGainMapSteerSpeedMax}{$-$120}

\newcommand{\SigGainMapSteerSpeedMin}{$-$234}

\newcommand{\SigNDrivers}{31}

\newcommand{\SigSetMapSOneMed}{339}
\newcommand{\SigSetMapSOneN}{407}

\newcommand{\SigSetMapSThreeMed}{49}

\newcommand{\SigUseFlagsPctMax}{$-$1.2}
\newcommand{\SigUseFlagsPctMin}{$-$2.4}

\newcommand{\SigUsePedalAllPedalPctMax}{$+$7.2}
\newcommand{\SigUsePedalAllPedalPctMin}{$+$4.2}

\newcommand{\SigUsePedalSpeedPedalPctMax}{$+$9.2}
\newcommand{\SigUsePedalSpeedPedalPctMin}{$-$0.5}
\newcommand{\SigUseSteerRatioMax}{3.8}
\newcommand{\SigUseSteerRatioMin}{3.4}

\newcommand{\SuppBeamMed}{46}

\newcommand{\SuppBeamNoneMed}{763}

\newcommand{\SuppBeamRoute}{61}

\newcommand{\SuppDrMed}{117}

\newcommand{\SuppDrivers}{12}

\newcommand{\SuppDrives}{27}

\newcommand{\SuppEpHpMed}{337}

\newcommand{\SuppMacroHi}{24}
\newcommand{\SuppMacroLo}{14}
\newcommand{\SuppMacroMed}{19}

\newcommand{\SuppMacroRoute}{85}

\newcommand{\SuppN}{46}

\newcommand{\SuppOpdTwoMed}{25}

\newcommand{\SuppOpdTwoRoute}{65}

\newcommand{\SuppPfHi}{492}
\newcommand{\SuppPfLo}{46}
\newcommand{\SuppPfMed}{75}

\newcommand{\SuppPfRoute}{59}

\newcommand{\TaxMapBeamNoneHmTestFiftyNoCandPct}{58}

\newcommand{\TaxMapBeamNoneHmTestFiftyRankMissPct}{24}
\newcommand{\TaxMapBeamNoneHmTestFiftyTopOneOkPct}{18}

\newcommand{\TaxMapBeamSteerHpTestFiftyNoCandPct}{28}

\newcommand{\TaxMapBeamSteerHpTestFiftyRankMissPct}{10}
\newcommand{\TaxMapBeamSteerHpTestFiftyTopOneOkPct}{61}

\newcommand{\TopkDewriHmTestBestFirst}{39}

\newcommand{\TopkDewriHmTestKOne}{293}
\newcommand{\TopkDewriHmTestKTen}{49}

\newcommand{\TopkElasticPathingHmTestKOne}{129}
\newcommand{\TopkElasticPathingHmTestKTen}{96}

\newcommand{\TopkElasticPathingHmTestOut}{1{,}855}
\newcommand{\TopkElasticPathingHmTestSingle}{1{,}484}

\newcommand{\TopkMapBeamNoneHmTestBestFirst}{21}

\newcommand{\TopkMapBeamNoneHmTestKOne}{1{,}314}
\newcommand{\TopkMapBeamNoneHmTestKTen}{72}
\newcommand{\TopkMapBeamNoneHmTestKThree}{210}

\newcommand{\TopkMapBeamSteerHmTestBestFirst}{54}

\newcommand{\TopkMapBeamSteerHmTestKOne}{41}
\newcommand{\TopkMapBeamSteerHmTestKTen}{27}

%% file: sections/abstract.tex
Open driving releases and usage-based insurance programs often claim to prevent privacy breaches by omitting GPS records when sharing or collecting vehicle data.
However, removing GPS coordinates from driving logs does not erase the location information carried by vehicle signals.
While prior route-inference attacks have demonstrated that speed, steering, and driving events can be matched to road networks to reconstruct trajectories, existing studies rely on inconsistent prior assumptions and evaluation metrics and are tested only on small, restricted datasets, making their results difficult to compare and their real-world effectiveness uncertain.
We release CANTraj, a naturalistic driving dataset pairing decoded CAN events with time-aligned GNSS trajectories (\NReleaseDrives{} drives, \NReleaseDrivers{} drivers), and CANTraj-Bench for evaluating trajectory recovery (\NDrives{} drives, \NDrivers{} drivers, \NHours{} hours, \NKm{}\,km).
On \NHeldoutDrives{} evaluation drives from \NHeldoutDrivers{} unseen drivers, we compare six re-implemented attacks and benchmark baselines under explicit information-access protocols, paired comparisons, and metrics incorporating failures.
Experiments show that combining steering signals with road-network constraints yields the largest gain: With a known start and heading, adding steering to speed raises the online map filter's five-minute route recovery from \ResPfNoneFiveTestRoute{}\,\% to \ResPfFiveTestRoute{}\,\%, with a median average displacement error of \ResPfFiveTestMed{}\,m versus \ResDrFiveTestMed{}\,m for map-free dead reckoning.
Offline beam search recovers \ResBeamFiveTestRoute{}\,\% of test routes with the heading and \ResBeamHmFiveTestRoute{}\,\% without it.
Because methods fail on different windows, we also evaluate an LLM harness that selects among candidate routes on \AgentCohortAllN{} five-minute windows; its best configuration recovers \AgentMaxVoteTestRoute{}\,\% of routes on test and \AgentMaxVoteFewShotRoute{}\,\% on few-shot, outperforming the best baseline on test and matching it on few-shot.
Data and code are available at \href{https://huggingface.co/datasets/HenryYHW/CANTraj}{Hugging Face} and \href{https://github.com/WangYuHang-cmd/GPS--CAN}{GitHub}.

%% file: sections/introduction.tex
\section{Introduction}\label{sec:intro}

Modern cars broadcast wheel speed, steering angle, pedal positions and yaw rate over the controller area network
(CAN) bus, which can be accessed through the on-board diagnostics (OBD) port. Usage-based insurance, fleet
telematics, aftermarket driver assistance and public driving datasets collect these signals, as do studies of
production driver-assistance systems~\cite{wang2025openlka,wang2026adasto}. Removing global navigation satellite
system (GNSS) coordinates may appear to protect location privacy, but the remaining vehicle signals can still reveal
where a car travelled~\cite{dewri2013inferring,gao2014elastic}.

Existing attacks exploit speed patterns, driving events or steering geometry. Speed-based methods match stops,
travelled distances and feasible turns to a road network, using path search, hidden Markov models or particle
filters~\cite{dewri2013inferring,gao2014elastic,zhou2017insurance,wahlstrom2016speed}.
Event-based methods derive routes from distances and turn directions~\cite{waltereit2019route}, learn manoeuvres
from brake signals (BBLT)~\cite{sarker2020bblt,sarker2023brake}, or match events detected from speed and accelerator
pedal position to road intersections (CAN-Trace)~\cite{lin2025cantrace}. Steering-based attacks either match the
steering profile to road curvature (SPy)~\cite{pese2020spy} or combine a vehicle motion model with map matching
(MacroTracking and OPD-II)~\cite{gazdag2023privacy,bianchi2025opd}. Their outputs range from ranked destinations and
candidate routes to position estimates along a trip.

These studies establish a privacy risk, but their results remain difficult to compare across signals and methods.
Their datasets differ in coverage and signal source: SPy evaluates 58 trips from five vehicle models, OPD-II reports
59 tracks from four cars, and Elastic Pathing uses 945 traces from two regions, including GPS-derived
speeds~\cite{pese2020spy,bianchi2025opd,gao2014elastic,gao2017elastic}. Comparing these published scores mixes differences in
signals, algorithms, drivers, vehicles and road networks. The assumed knowledge also varies: a known start and
heading, a containing area, or additional traffic information~\cite{bianchi2025opd,pese2020spy,lin2025cantrace,zhou2019privacy}.
Reported success can mean a highly ranked destination, trajectory overlap, or nodes covered by the union of several
candidate routes~\cite{dewri2013inferring,bianchi2025opd,lin2025cantrace}; these scores do not measure the same recovery
task. Individual sensitivity studies leave open how methods compare on the same drives, how much each signal adds
with the reconstruction model fixed, and how accuracy, failures and cost change with the available information and
trip duration. Data owners need these comparisons to assess a CAN release; localisation systems need them when GNSS
is unavailable.

We address these gaps through CANTraj and CANTraj-Bench, with three research questions. \textbf{RQ1:} How accurately
can existing methods recover trajectories, and which CAN signals improve recovery? \textbf{RQ2:} How do initial
heading, online or offline access, and trip duration affect accuracy? \textbf{RQ3:} What explains failures, and how
do accuracy, computational cost and resistance to degraded signals trade off?

Our contributions are threefold.

(I) \textbf{Dataset and benchmark.} We introduce CANTraj, a naturalistic
driving dataset of \NReleaseDrives{} drives from \NReleaseDrivers{} drivers
with decoded CAN signals and time-aligned GNSS trajectories. CANTraj-Bench
uses \NDrives{} of these drives, spanning \NHours{} hours and \NKm{}\,km,
to evaluate online and offline trajectory recovery from different
combinations of CAN signals. It defines \NEligibleInFolds{} evaluation
episodes and specifies signal sources, map access, and failure scoring rule (Sections~\ref{sec:problem} and~\ref{sec:data}).

(II) \textbf{Unified cross-method comparison.} We re-implement six published
attacks and compare them with dead reckoning, map-based particle filtering,
and offline beam search. Where available, we use the original authors'
sample data to check our implementations. The main evaluation covers
\NHeldoutDrives{} drives from \NHeldoutDrivers{} drivers across the test
and few-shot splits, with paired comparisons on common episodes. We
document each method's adaptations and information requirements
(Section~\ref{sec:methods}) and separately evaluate an LLM harness that
selects among candidate routes.

(III) \textbf{Recovery and failure analysis.} Steering provides the largest
gain among the CAN signals tested. With speed and steering, the median
error on five-minute test windows is \ResPfFiveTestMed{}\,m for online map
filtering and \ResDrFiveTestMed{}\,m for map-free dead reckoning. Offline
beam search is less sensitive to missing initial heading than online
filtering and avoids some of its largest errors on longer windows.
Different methods recover different windows, and quantising steering to
10$^\circ$ or speed to 1\,m/s leaves the tested map-based attacks largely
effective (Section~\ref{sec:results}).

%% file: sections/related.tex
\section{Related Work}\label{sec:related}

\paragraph{Route inference from vehicle signals}
Early speed-based attacks search for destinations or paths consistent with stops, distances and feasible
turns~\cite{dewri2013inferring,gao2014elastic,gao2017elastic}. Wahlstr\"om et al.~\cite{wahlstrom2016speed} use a map
particle filter with lateral-force and speed-limit constraints. Zhou et al.~\cite{zhou2017insurance,zhou2019privacy}
combine hidden Markov search and speed-sequence matching with traffic information. These studies include
GPS-derived speed traces as well as OBD measurements, and the later Elastic Pathing and Zhou papers reuse earlier
datasets. Their results demonstrate leakage from speed, but do not isolate the effect of signal source or decoder
on a common vehicle population.

Richer signals expose additional route structure. BBLT and its journal extension infer manoeuvres from brake
signals before searching the road network~\cite{sarker2020bblt,sarker2023brake}; CAN-Trace uses speed and pedal
events~\cite{lin2025cantrace}. Waltereit et al.~\cite{waltereit2019route} match distance and turn sequences, which can
come from CAN or phone sensors, but evaluate GPS-derived events with injected errors. SPy matches steering to road
curvature~\cite{pese2020spy}; MacroTracking and OPD-II integrate speed and steering and correct the resulting path
with a map~\cite{gazdag2023privacy,bianchi2025opd}. Their evaluations answer different questions: MacroTracking
measures proximity to a reference path, OPD-II reports trajectory similarity, and CAN-Trace includes coverage by
multiple candidates. We retain route and position metrics separately and count unsuccessful searches, so unsuccessful searches remain visible in the comparison.

\textbf{Vehicle localisation and learned motion models}
Map matching traditionally assigns noisy positions to roads~\cite{newson2009hidden}, while sparse trajectory
recovery also receives intermittent GPS positions~\cite{tian2025trmma}. Arc-length map matching uses vehicle sensors
with a known lane centreline and recent GPS initialisation~\cite{javed2024arclength}. These inputs differ from a CAN
release with only one known position. Learned navigation offers complementary designs: AI-IMU adapts a filter's
measurement noise~\cite{brossard2020aiimu}; FormerReckoning and RoTrax learn from inertial
sequences~\cite{li2024formerreckoning,pandey2026rotrax}; S-TRAVERSE combines vehicle speed with a smartphone
IMU~\cite{jun2025straverse}. GHNet estimates heading from GNSS velocity~\cite{dahan2023ghnet}, and rapid alignment
methods use a GNSS trajectory window~\cite{chen2023heading,wu2024heading}. They motivate learned motion and heading
studies, but their additional sensors or GNSS observations must be distinguished from CAN-only inputs. Our signal source labels make this distinction explicit.

\textbf{Location protection and evaluation data}
RAoPT and VehiTrack reconstruct trajectories from perturbed locations~\cite{buchholz2022raopt,yadav2024transprotect}.
TransProtect generates plausible substitute locations, while Olympus and adversarial mobility prediction learn
representations that balance an authorised task with sensitive inference~\cite{yadav2024transprotect,raval2019olympus,zhan2022mobility}.
These studies motivate testing adaptive attacks and assessing downstream utility. In our experiments, map-based attacks using their original settings still recover routes from quantised CAN signals.
Likewise, accuracy and computational cost should be reported together, as emphasised in trajectory-similarity evaluation~\cite{chang2024trajsimi}. 
Public driving datasets serve different purposes: comma2k19 provides CAN and GNSS for a highway commute~\cite{schafer2018commute}, while DriveDNA
provides naturalistic data without GPS coordinates~\cite{wang2026drivedna}. CANTraj adds paired CAN and GNSS
data across hundreds of drivers and vehicle platforms, together with a common evaluation of six attacks and controlled signal comparisons.

%% file: sections/problem.tex
\section{Trajectory Recovery Task and Evaluation}
\label{sec:problem}

\paragraph{Task}
We study how accurately a vehicle can be tracked from CAN signals over different time windows when its starting position is known.
Given the position $p_0$ at time $t_0$, each method uses the available vehicle signals to estimate the trajectory over $[t_0,t_0+D]$.
We query the method at the timestamps of recorded GPS fixes, with at most one fix per second and no interpolation.
The method must return a position estimate at each query time.
If a method outputs only a route, we integrate the speed used by that method to obtain the distance travelled at each query time, then use that distance to locate the vehicle along the predicted route.
For SPy~\cite{pese2020spy}, we use the time alignment from its own warping path.
In this conversion, the GPS reference provides only the query timestamps.

\paragraph{Priors and protocols}
Three priors are switched independently, giving eight protocols written \textsc{O/F}-\textsc{N/M}-\textsc{H+/H$-$}.
\emph{Online} (O): an event becomes visible only after its logged time, the episode length is unknown, and the method is
queried as the reference seconds arrive; \emph{offline} (F): all events of the episode are visible, nothing outside it.
\emph{Map} (M): an OpenStreetMap (OSM) drive network; the usable region is the disc of radius
$5\,\mathrm{km} + 40\,\mathrm{m/s}\cdot D$ (at most 60\,km) around $p_0$, and only that disc is visible to the
benchmark's map methods (Section~\ref{sec:data}); OPD-II's Meili matcher searches
its Valhalla region tiles within 200\,m of the dead-reckoned trace and is not cut to the disc (Table~\ref{tab:methods}).
\emph{Heading} (H+): the GPS course at $t_0$ is given; under H$-$ it is withheld. The start position is always given,
except to the two area adaptations, which receive a 5\,km square centred on it (Section~\ref{sec:methods}; the method
does not read $p_0$, but the square's placement uses it). This study reports five of the eight protocols: without the
map only \textsc{O}-\textsc{N}-\textsc{H+}, with the map all four. Published methods are run under the protocol that
matches their own assumptions and labelled with their look-ahead (OPD-II decides a position up to 3\,s later).

\paragraph{Inputs}
Each method receives only the columns allowed by its signal set, and static vehicle
parameters (wheelbase, steering ratio) come from the drive's own logged car parameters unless a variant is named
``calibrated'' (training-split calibration). Every signal is labelled by its source: C0 read from the vehicle bus
(speed, steering angle, pedals, yaw rate), C1 derived from CAN inside the vehicle software, I0 the logger's raw
inertial measurement unit (IMU), and F
produced by the driving software's localizer, which used GPS in the software versions of this data. Class F results
are reported only as a reference and never as CAN-only. Beyond the permitted start position and initial heading, GPS positions, speed, course, accuracy fields and absolute time are excluded.

\paragraph{Episodes}
An anchor $t_0$ is a GPS fix with speed $\ge 3$\,m/s, course accuracy $\le 5^\circ$ (and horizontal accuracy $<10$\,m
for the u-blox receiver) and CAN speed $\ge 2.5$\,m/s; $p_0$ is the fix's position. Windows of 1, 5 and 15 minutes and
the rest of the drive are cut from anchors; an episode is eligible when at least 80\,\% of its seconds have a fix,
no reference gap exceeds 30\,s and the path is at least 100\,m long. Windows of different lengths overlap and come from
different drives, so the horizon study (RQ2) uses a fixed cohort with one start per drive.

\paragraph{Metrics}
For an episode with errors $e_t$ between prediction and reference: average displacement error (ADE, mean of $e_t$);
coverage $c_r$ (share of seconds with $e_t\le r$, $r=100$\,m unless stated); time-free path recall and precision at
50\,m (share of the reference path within 50\,m of the predicted path and vice versa, both resampled at 5\,m); route
recovered $=$ path recall@50 $\ge 0.8$, which says the route was followed, not that every second was right; the
end-of-window error (reported as the share beyond 500\,m). Two failure definitions are named separately: \emph{tail-lost} (the final fifth of the
window stays beyond 100\,m) and \emph{terminal excursion} (the window ends beyond 100\,m; its onset is the first second
after the last return within 100\,m). The text reports ADE, $c_{100}$, route recovered and the two failure shares; the
released tables carry every metric.

\paragraph{Failures and statistics}
The episodes a method is expected to answer are fixed before looking at its outputs (eligible episodes of the window,
start on the map for map protocols, the method's signals valid on the drive, a start heading if the method needs one).
An expected episode without output (time-out, no match, error) is a failure: its ADE is $+\infty$ in medians and
distributions and its coverage and route flags are zero; means are reported only on episodes where every compared
method has an output, next to the failure rate. Confidence intervals (CIs) are percentile bootstraps that resample
\emph{drivers} (one driver can own hundreds of drives; 1{,}000 resamples, 500 for the cohort and signal studies and 300
for the top-$K$ analysis); a driver-equal median summarises each driver first. Comparisons
between methods or signal sets are paired on common episodes: paired median and mean differences use the episodes where
both have an output, and the share of episodes where one method is closer counts its failures as losses (episodes where
both fail are left out). A paired interval that includes zero means that this sample gives no clear evidence either
way, not that the two are equivalent. Five comparisons are primary: the map filter against the beam (steering, heading
known), speed against speed and steering (signal study), heading known against unknown (beam and filter), OPD-II against
the filter, and learned against nominal motion sources; every other comparison is exploratory, and its interval describes
uncertainty without a correction for multiplicity. For learned methods every statistic, bootstrap bounds included, is the mean over three
training seeds, with the SD over seeds. 

%% file: sections/data.tex
\section{Data and Quality}\label{sec:data}

\begin{figure}[htbp]
    \centering
    \includegraphics[width=0.49\textwidth]{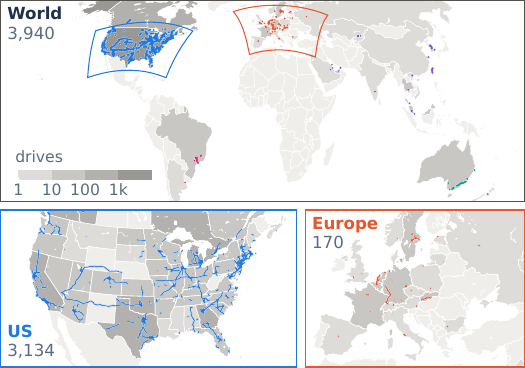}
    \caption{Geographic distribution of the CANTraj drives. Gray gives the number of drives per country, and per state or province in the US panel. Track colour gives the continent. Tags give each region's drive count. One area is not drawn, and its drives are not counted in the gray levels.}
    \label{fig:dataset_overview}
\end{figure}

\paragraph{Corpus}
CANTraj follows the collection and processing similar to DriveDNA~\cite{wang2026drivedna}: naturalistic drives logged by aftermarket driver-assistance systems on the drivers' own cars, with the raw logs re-decoded, cleaned and pseudonymised;
unlike that release, which carries no GPS coordinates, it pairs CAN and GNSS. CANTraj-Bench uses \NDrives{} drives of \NDrivers{} drivers, \NPlatforms{} vehicle platforms of \NBrands{} brands, \NHours{} hours and \NKm{}\,km: every raw log re-decoded into time-stamped message tables (100\,Hz where the full log was kept, 10\,Hz for drives with only the compressed log); the drive's GNSS track on the same time axis, with the first and last two minutes of every drive cut; quality metadata and the splits below (pseudonymous drivers, no absolute time, no raw identifiers, Fig~\ref{fig:dataset_overview}). 
The release holds \NReleaseDrives{} drives of \NReleaseDrivers{} drivers (\NReleaseHours{} hours of GNSS track): the drives above and \NReleaseUnusedDrives{} drives that passed the same GNSS checks but yield no eligible episode (Section~\ref{sec:problem}): \NReleaseNoAnchorDrives{} without an admissible anchor, \NReleaseNoWindowDrives{} whose windows all have GNSS coverage below 80\,\%, and \NReleasePlatformDrives{} whose vehicle platform is unconfirmed; they stay in the release for other windowing rules and are not used here. 
The drives' map discs touch OpenStreetMap extract regions in \NRegionUS{} US states, \NRegionCA{} Canadian provinces and \NRegionOther{} further regions in Europe,
Asia, Oceania and South America; North America dominates. Drivers are split into disjoint folds (train, validation, test, few-shot
evaluation; \NUnassignedDrives{} drives of \NUnassignedDrivers{} drivers belong to no fold and are not used); all reported results are on test (\NTestDrivers{} drivers, \NTestDrives{} drives) and few-shot
(\NFewShotDrivers{} drivers, \NFewShotDrives{} drives), whose drivers never occur in training.

\input{tables/methods.tex}

\paragraph{GNSS Reference}
The reference is the GNSS track logged with each drive, one real fix per second: a modem receiver on \NDrivesQcom{} drives (about
1\,Hz) and a u-blox receiver on \NDrivesUblox{} drives (up to 10\,Hz, reduced to 1\,Hz). GNSS time lags the bus by a
receiver-dependent delay, estimated on the training fold by cross-correlating GNSS and CAN speed (0.7\,s modem,
0.2\,s u-blox) and applied to the reference for every method alike. Eligibility (Section~\ref{sec:problem}) leaves
\NTestFiveMin{} test and \NFewShotFiveMin{} few-shot 5-minute episodes, \NTestFifteenMin{} and \NFewShotFifteenMin{} 15-minute episodes, and \NTestFullMin{} and
\NFewShotFullMin{} whole-drive episodes. Drives whose vehicle platform could not be confirmed from the logged
fingerprint are excluded from the CAN-only board (\NPlatformUnconfirmedDrives{} drives of \NPlatformUnconfirmedDrivers{}
drivers).

\paragraph{Signals}
Speed and steering angle are present on almost every drive. The accelerator pedal position is a continuous signal on
only part of the brands (\NPedalTestDrives{} of \NTestDrives{} test and \NPedalFewShotDrives{} of \NFewShotDrives{}
few-shot drives); where present it correlates with the longitudinal acceleration, at a brand-specific scale. The brake
field mixes pedal position, pressure and torque across brands; the pedal-pressed flags are available almost everywhere. Methods that need the accelerator (CAN-Trace) or continuous pedals (the signal sets S2c and S4c of
Section~\ref{sec:methods}) are therefore
evaluated on the pedal-valid subset, and the pressed flags are a separately named set (S2b, S4b), not a substitute.

\paragraph{Maps}
Road graphs are cut from Geofabrik OpenStreetMap extracts of one snapshot into 0.25$^\circ$ tiles (drivable highway
classes, one-way and turn restrictions). An audit of the first map build against its sources found \NTilesTruncated{} of the
\NTilesVtwo{} needed tiles cut incompletely at the margins of early tiling passes (starts there had no road nearby, and
map methods fell back to an arbitrary edge); every tile was re-cut from all intersecting extracts (the rebuilt road network), used by every
map-based number here. With the rebuilt road network the median distance from the start to the nearest drivable road is \MapStartDistMedian\,m; the
\MapStartMissingMain{} main-fold episodes without any road within 400\,m are excluded from the map protocols for all
methods, by a rule that looks at the map and the start only. Within one drive a single cached graph serves all episodes,
but each episode sees only the tiles of its own disc, with edge bearings in its local frame; exposing the whole
per-drive graph changes no output (\AThreeWindows{} test windows of \AThreeDrives{} drives: \AThreeIdenticalPct{}\,\%
identical within 1\,m; a changed filter seed moves \AThreeSeedMovedPctFive{}\,\%).

%% file: tables/methods.tex
\begin{table*}[!tp]
\centering
\caption{Methods and evaluation settings. $p_0$ and $\theta_0$ denote the start position and heading. ``OSM disc'' is each episode's available map area. DR, PF, and GRU stand for dead reckoning, particle filter, and gated recurrent unit. S1--S4 are the signal sets in Section~\ref{sec:methods}.}
\label{tab:methods}
\footnotesize
\setlength{\tabcolsep}{5pt}
\renewcommand{\arraystretch}{1.07}
\setlength{\aboverulesep}{0pt}\setlength{\belowrulesep}{0pt}
\begin{threeparttable}
\begin{tabularx}{\textwidth}{L{1.36} c c L{0.73} c c c L{0.92} L{1.27} L{0.72}}
\rowcolor{cHead}
 & \multicolumn{3}{l}{\hdspan{Signals}} & \multicolumn{2}{l}{\hdspan{Priors}} & & & & \\
\rowcolor{cHead}
\hd{Method} & \hd{Speed} & \hd{Steer} & \hd{Other} & \hd{$p_0$} & \hd{$\theta_0$} & \hd{Map} & \hd{Mode} & \hd{Training data} & \hd{Output} \\
\grp{cBase}{Map-free baselines}
             Hold start        & \N & \N & \N & \Y & \N & \N & \On & \N & position \\
\tint{cBase} Straight line     & \Y & \N & \N & \Y & \Y & \N & \On & \N & position \\
             Steering DR       & \Y & \Y & \N & \Y & \Y & \N & \On & \N{} / calib.\tnote{a} & position \\
\tint{cBase} Start prior       & \Y & \N & \N & \Y & \Y & \N & \On & train GPS (statistics) & position \\
             Route retrieval   & \Y & \N & \N & \Y & \Y & \N & \On & train GPS (routes) & position \\
\grp{cLearn}{Learned motion models}
              Learned GRU               & \Y & \Y   & \N           & \Y & \Y & \Opt & \On & train GPS & position \\
\tint{cLearn} Learned physics residual  & \Y & \Y   & \N           & \Y & \Y & \Opt & \On & train GPS & position \\
              Signal-set GRU (S1 to S4)   & \Y & \Opt & pedal (\Opt) & \Y & \Y & \Opt & \On & train GPS & position \\
\grp{cMap}{Map baselines}
            Map PF   & \Y & \Opt & \N & \Y & \Opt & \Y & \On  & \N & position \\
\tint{cMap} Map beam & \Y & \Opt & \N & \Y & \Opt & \Y & \Off & \N & top-10 paths \\
\grp{cPub}{Published attacks}
            MacroTracking~\cite{gazdag2023privacy} & \Y & \Y & \N & \Y & \Y & \Y & \On & \N{} / calib.\tnote{a} & position \\
\tint{cPub} OPD-II~\cite{bianchi2025opd}                 & \Y & \Y & \N & \Y & \Y & Meili\tnote{e} & \Del\tnote{b} & \N{} / calib.\tnote{a} & position \\
            Elastic Pathing~\cite{gao2017elastic}     & \Y & \N & \N & \Y\rlap{\tnote{c}} & \Opt & \Y & \Off & val (2 values) & top-$n$ paths \\
\tint{cPub} Dewri et al.~\cite{dewri2013inferring}          & \Y & \N & \N & \Y\rlap{\tnote{c}} & \Opt & \Y & \Off & \N & top-10 paths \\
            CAN-Trace~\cite{lin2025cantrace}          & \Y & \N & accel.\ pedal & area\tnote{c}\,/\,$p_0$ & \N & \Y & \Off & \N & top-10 paths \\
\tint{cPub} SPy~\cite{pese2020spy}                     & \N & \Y & \N & area\tnote{c}\,/\,$p_0$ & \Opt & \Y & \Off & val (2 values) & path \\
\grp{cRef}{Reference (GPS-informed, not CAN-only)}
            Curvature DR (class F) & \Y & \N & localizer curv. & \Y & \Y & \Opt & \On & GPS-informed\tnote{d} & position \\
\arrayrulecolor{cHead}\specialrule{1pt}{0pt}{0pt}\arrayrulecolor{black}
\end{tabularx}
\begin{tablenotes}[flushleft]\scriptsize
\item[] \Y: used; \Opt: optional or protocol-dependent; \N: not used.
\item[] $^{\rm a}$\,Variant with training-fold steering calibration.
$^{\rm b}$\,Each position is fixed when its 3\,s batch is map-matched; a strictly online read-out is also reported.
$^{\rm c}$\,Native prior first, then our adaptations to the shared protocols; ``area'' is a 5\,km square centred on the known start.
Elastic Pathing and Dewri et al.\ read speed at 1\,Hz.
$^{\rm d}$\,The localizer used GPS in the software versions of this data; reference only.
$^{\rm e}$\,Meili matches OSM roads within its Valhalla region tiles, 200\,m around the dead-reckoned trace, and is not cut to the episode's disc.
\end{tablenotes}
\end{threeparttable}
\end{table*}

%% file: sections/methods.tex
\section{Methods and Setup}\label{sec:methods}

\begin{figure*}[!tp]\centering
\includegraphics[width=\textwidth,trim=8 4 6 0,clip]{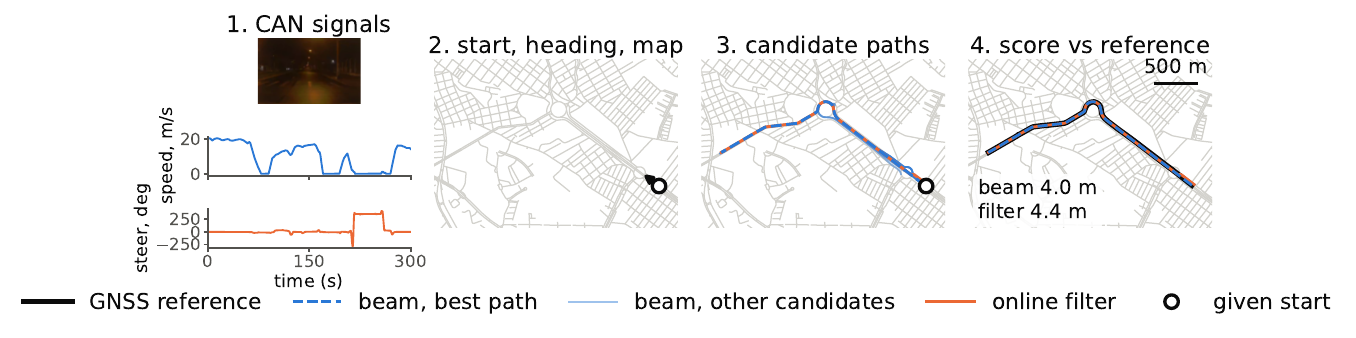}
\caption{Reconstruction and evaluation of a five-minute window: a method sees the CAN signals, the start, the heading when the
protocol gives it and the roads of the visible disc; the offline beam keeps ten candidate paths, the online filter one;
every method is scored at the reference seconds. The camera frame shows the scene and is not an input.}\label{fig:method}
\end{figure*}
Table~\ref{tab:methods} lists every method with its signals, priors, look-ahead and use of training data; Fig.~\ref{fig:method}
shows the pipeline on one window. All methods
share one replay engine: online methods receive the messages up to the query time, offline methods the whole episode,
and the reference loader is disabled while any method runs.

\paragraph{Published methods}
We re-implemented six published attacks under the benchmark's interface. Among these six attacks, we obtained OPD-II evaluation logs and a
MacroTracking sample trace; we use them to check our transcription against the released code, not to
recompute the papers' scores.
\emph{MacroTracking}~\cite{gazdag2023privacy} (speed, steering; online): 1\,s dead-reckoning updates and, every 20\,m,
a pull toward the nearest road edge weighted by the distance to the next intersection. The released code works in Web
Mercator units with the speed scaled by 1.1 (constants fitted to the authors' car and city); we use metres and the
logged speed. With the authors' constants on their sample trace, the re-implementation stays within a median of
\MacroSampleMedian\,m of their committed trajectory.
\emph{OPD-II}~\cite{bianchi2025opd} (speed, steering): bicycle-model dead reckoning in 0.1\,s windows, re-snapped every
30 windows by Valhalla/Meili map matching~\cite{newson2009hidden} on the same road network; its released tool outputs
each position once matched (at most 3\,s later), and we also report a strictly online read-out. On \OpdSampleLogs{} of the
authors' 59 sample logs, the re-implementation and the unmodified released script, run on the same decoded inputs and
tiles, differ by a median of at most \OpdSampleMaxMedian\,m. The authors' per-car ``wheelbase'' values (3.85 to 4.9\,m) and
angle scaling are a car-specific calibration of the bicycle model; we use the physical model with each drive's own
wheelbase and steering ratio (a training-calibrated variant changes the paired median by \PairOpdCalVsOpdFiveTestMed{}\,m)
and follow the code where it differs from the paper.
\emph{Elastic Pathing}~\cite{gao2014elastic,gao2017elastic} (speed only; offline): best-first search over road paths
that stretches or compresses the speed-integrated distance at stops and turns, with turn feasibility from road
geometry and speed limits; we follow the 2017 version. Values the paper leaves
open (side friction, the speed counted as a stop) were chosen on validation drives only.
\emph{CAN-Trace}~\cite{lin2025cantrace} (speed and accelerator pedal; offline): stops and pedal releases become nodes of
a line graph whose edge lengths come from integrated speed; the graph is matched to the intersection graph of the map
by subgraph isomorphism with a growing tolerance. Its original prior is an area containing the trip, not a start
point. We report a known-start adaptation and an area adaptation whose road graph is cut to a 5\,km square centred on
the known start: the method does not read $p_0$, but the square's placement uses it, and a window whose route leaves
the square cannot be matched (Section~\ref{sec:rq1}). Its open values (the number of K-means clusters, the stop
threshold of 0.05\,m/s and the 1\,\% pedal-minimum band) were fixed without tuning.
\emph{Dewri et al.}~\cite{dewri2013inferring} (speed only; offline): a depth-first search from the known start keeps
the road paths that pass an intersection at every stop of the trip, stay within 0.1\,mile of a shortest path, turn only
where the speed is below 25\,mph and end at the trip length; candidates are ranked by the likelihood of the observed
speeds under a speed-limit model that slows down at turns and stops. The paper ranks destinations; we lay the distance
travelled along each of the top ten paths. No value was tuned; its search budget is Elastic Pathing's, counted in node
expansions and shared by the start edges.
\emph{SPy}~\cite{pese2020spy} (steering only, by design; offline): the z-scored steering profile between turns is
matched by dynamic time warping to the curvature of every run of road segments in an area (RoCuMa), the most curved
section first, and its neighbours are chained by the turn direction. The paper uses no speed and no start point and
searches a city-wide road database; we report an area adaptation (a 5\,km square centred on the known start, our size, as for
CAN-Trace) and a known-start adaptation, with the two values the paper leaves open (the moving-average window and the
steering level that ends a turn) chosen on validation drives; the paper's 90$^\circ$ turn threshold is kept, and the
search budget is $10^8$ dynamic-time-warping cells per query. Our tiles carry no road names, so a road is a chain of ways joined at their straightest
continuation; positions per second come from the method's own warping path (our extension). BBLT~\cite{sarker2020bblt,sarker2023brake} and Zhou et al.~\cite{zhou2017insurance,zhou2019privacy} use additional traffic or road priors that we do not supply. Waltereit et al.~\cite{waltereit2019route} accept distance and turn events, potentially from CAN; their end-to-end event extraction is outside our six re-implementations. These methods remain related work. Deviations from the released code
are listed in the artifact; two bugs in released code are kept in the main rows
because the published results come from that code: MacroTracking compares headings in raw degrees without
wrapping at north, and OPD-II writes $360-b$ instead of $360+b$ for a negative bearing $b$. The as-released angle
handling is the MacroTracking variant that stays closest to the authors' committed trajectory. In sensitivity runs (not
in Table~\ref{tab:main}), fixing OPD-II's bug changes
nothing measurable (paired mean difference \PairOpdFixFifteenTestMean{}\,m [\PairOpdFixFifteenTestMeanLo{},
\PairOpdFixFifteenTestMeanHi{}] at fifteen minutes on test); fixing MacroTracking's leaves the paired median unchanged
(\PairMacroFixFifteenTestMed{}\,m) but removes some large failures (paired mean \PairMacroFixFifteenTestMean{}\,m
[\PairMacroFixFifteenTestMeanLo{}, \PairMacroFixFifteenTestMeanHi{}]).

\paragraph{Benchmark baselines}
Dead reckoning (DR; a straight line from speed, or the bicycle model from steering with the logged car parameters, with
an online steering-offset estimate, or with a training-fold calibration); holding the start; a start prior and an
identity-free route retrieval from training drives; a map particle filter (PF; 2,000 particles on road edges, steering-
or speed-driven, online weighted-mean output) and an offline beam search (Algorithm~\ref{alg:beam}; $B{=}200$, $K{=}10$,
$\sigma_\mathrm{h}{=}15^\circ$, $\sigma_\mathrm{turn}{=}20^\circ$, $w_\mathrm{u}{=}0.05$, $w_\mathrm{d}{=}0.1$,
$\gamma{=}0.3$, $T{=}1$; a hypothesis turning near the next node may cross it early; the speed-only variant takes
successors uniformly, forbids U-turns and uses the filter's speed and road-class penalty instead of the heading term). A lineage
read-out of the same particle filter (the ancestry of the heaviest final particle) separates the effect of the
decoder's output from its motion model.

\begin{algorithm}[t]
\small
\caption{Offline map beam search (benchmark offline decoder).}\label{alg:beam}
\begin{algorithmic}[1]
\Require Window events; start $p_0$; heading $\theta_0$ (or none); visible graph $G$ with edge bearings $b(e)$ and lengths $|e|$; beam width $B$; top-$K$
\Ensure Top-$K$ trajectories with positions and confidence scores
\State Integrate events to get distance $s(t_i)$ and dead-reckoned heading $\psi(t_i)$ at reference timestamps $t_1, \ldots, t_n$
\State Initialize $H \gets \{(e, o{=}0, \phi, S{=}0)\}$ for drivable edges $e$ within $50\,\mathrm{m}$ of $p_0$: \Comment{prior alignment}
  \Statex \hspace{\algorithmicindent} $\bullet$ \textcolor{AlgBlue}{\textbf{with $\theta_0$ ($\text{H}^+$):}} keep $e$ within $45^\circ$ of $\theta_0$, set $\phi \gets \theta_0 - \psi(t_1)$
  \Statex \hspace{\algorithmicindent} $\bullet$ \textcolor{AlgBlue}{\textbf{without $\theta_0$ ($\text{H}^-$):}} keep both directions, set $\phi \gets b(e) - \psi(t_1)$
\For{$i = 2 \ldots n$}
  \For{\textbf{each} hypothesis $(e, o, \phi, S) \in H$}
    \State $o \gets o + s(t_i) - s(t_{i-1})$;\quad $\theta \gets \psi(t_i) + \phi$ \Comment{advance along edge}
    \While{$o > |e|$} \Comment{cross node transition}
      \State \textbf{if} $e$ has no successor \textbf{then} $o \gets |e|$, $S \mathrel{+}= \log w_\mathrm{d}$, \textbf{break}
      \State Branch to each successor $e'$ with $o \gets o - |e|$, $e \gets e'$, and
      \Statex \hspace{2\algorithmicindent} $S \mathrel{+}= \log \mathrm{softmax}_{e'}\!\Big({-\tfrac{1}{2}}\Big(\frac{b(e') - \theta}{\sigma_\mathrm{turn}}\Big)^{\!2} + \mathbb{I}(e'\text{ reverses})\log w_\mathrm{u}\Big)$
    \EndWhile
    \State $S \mathrel{-}= \tfrac{1}{2}\big((\theta - b(e))/\sigma_\mathrm{h}\big)^2$;\quad \algbox{$\phi \mathrel{-}= \gamma\,\mathrm{wrap}(\theta - b(e))$} \Comment{drift correction}
  \EndFor
  \State \textcolor{AlgBlue}{\textbf{Viterbi-merge}} hypotheses in $H$ sharing \algbox{$(e, \lfloor o / 10\,\mathrm{m} \rfloor)$} by keeping maximum $S$, then prune $H$ to top-$B$ \Comment{prevent beam collapse}
\EndFor
\State \Return Top-$K$ distinct edge sequences back-traced from $H$, interpolated positions from $o$, and normalized scores $\mathrm{softmax}(S_k / T)$
\end{algorithmic}
\end{algorithm}

\paragraph{Learned models and the signal study}
A causal gated recurrent unit (GRU; 64 units, platform and brand embeddings, 10\,Hz ticks) predicts the heading rate
and a speed correction from speed and steering, trained on training-fold GPS with model selection on validation chunks
(three seeds); it runs without a map and as the motion source of the particle filter. A second learned model, the
physics residual, keeps the bicycle model and learns per-platform gain, offset and speed-scale corrections plus a small
GRU residual, trained the same way. For the signal-combination study the same
architecture gets a second heading output, an unsigned turn magnitude, because without steering the \emph{direction}
of a turn is not identifiable while its occurrence is; one decoder (the particle filter with a per-particle turn
budget for the magnitude not explained by the signed rate) is used for every signal set. The six signal sets are S1
(speed), S2c and S2b (S1 plus the continuous accelerator pedal or the pressed flags), S3 (speed and steering) and S4c
and S4b (S3 plus the pedal or the flags); they differ only in which signal columns are non-zero and are trained on the
same drives with the same budget.

\paragraph{Setup}
All runs use one workstation (Intel Core i9-12900K, 24 threads, 31\,GB of memory; a consumer GPU trains the learned
models), Python~3.12 with NumPy, SciPy and PyTorch, and pyvalhalla~3.9.1 for OPD-II; the language model of
Section~\ref{sec:agent} is served from one H100 GPU on a separate host. Search methods have a per-episode
budget (Elastic Pathing: 0.2\,s of search time per second of the window, 30 to 300\,s, and 70 branch calls per second,
20{,}000 to 200{,}000; CAN-Trace: 120\,s and 200{,}000 expansions; Dewri et al.\ and SPy as stated above), and an
episode whose search ends without a result counts as a failure. Parameters of the benchmark baselines were fixed in an earlier version on training and
validation drives and not changed for this study.

%% file: sections/results.tex
\section{Results}\label{sec:results}

\begin{figure*}[htbp]\centering
\includegraphics[width=0.8\textwidth]{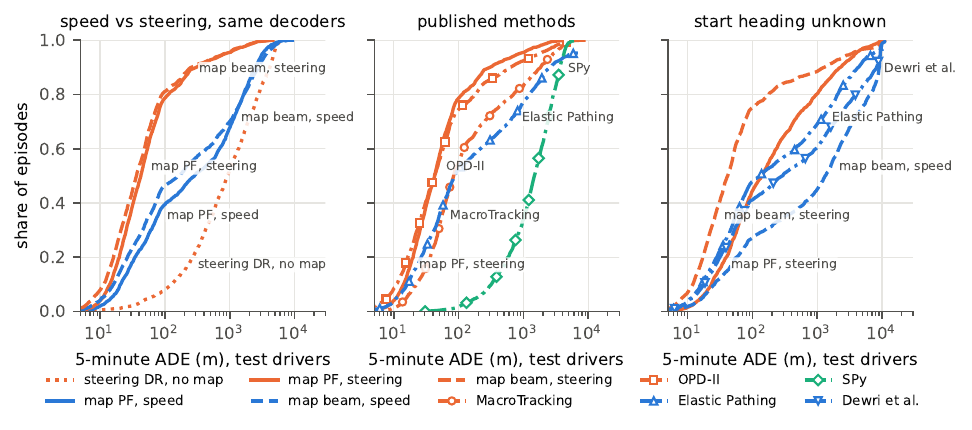}
\caption{Distribution of the five-minute ADE on the test drivers (\ResPfFiveTestN{} windows per method; expected but
unanswered windows stay in the denominator, so a curve ends at one minus the failure rate). Hue: signal set (blue
speed, orange speed and steering, green steering only); line style: decoder (solid online filter, dashed offline beam,
dotted no map, dash-dot with markers published methods). In the published-methods panel Elastic Pathing and SPy
use the start heading and OPD-II is its 3\,s delayed output; CAN-Trace is omitted (too few outputs,
Section~\ref{sec:rq1}).}
\label{fig:cdf}
\end{figure*}

\subsection{RQ1: Methods and signal contributions}\label{sec:rq1}
\input{tables/main_board.tex}
\begin{table*}[!htbp]
\centering\small
\setlength{\tabcolsep}{4pt}
\setlength{\aboverulesep}{0pt}
\setlength{\belowrulesep}{0pt}
\renewcommand{\arraystretch}{1.35}
\arrayrulecolor{TabRule}

\providecommand{\rqtag}[2]{%
  {\setlength{\fboxsep}{1.6pt}%
   \colorbox{#1!26}{%
     \textcolor{#1!78!black}{%
       \scriptsize\sffamily\bfseries\strut #2}}}%
}
\providecommand{\rqname}[2]{%
  \textbf{\textcolor{#1!78!black}{#2}}%
}

\caption{Main findings and the comparisons that support them.}
\label{tab:findings}

\begin{tabularx}{\textwidth}{
  c
  >{\raggedright\arraybackslash}p{58pt}
  >{\hsize=0.80\hsize\raggedright\arraybackslash}X
  >{\hsize=1.20\hsize\raggedright\arraybackslash}X
}
\toprule

\rowcolor{cHead}
\multicolumn{2}{l}{\hd{Question}}
& \hd{Finding}
& \hd{Evidence and scope} \\

\midrule

\tint{cMap}
\rqtag{cMap}{RQ1}
& \rqname{cMap}{methods and signals}
& The map and steering provide the largest gains; learned motion and pedals add little with the map.
& Five-minute test medians: steering DR \ResDrFiveTestMed{}\,m, map filter \ResPfFiveTestMed{}\,m. Signal comparisons fix the learner, decoder and drives. \\

\arrayrulecolor{white}
\specialrule{1.2pt}{0pt}{0pt}

\tint{cLearn}
\rqtag{cLearn}{RQ2}
& \rqname{cLearn}{information and duration}
& Offline decoding reduces sensitivity to initial heading and avoids some large tracking failures.
& Beam medians: \ResBeamFiveTestMed{}\,m with heading, \ResBeamHmFiveTestMed{}\,m without. Heading still changes route recovery from \ResBeamHmFiveTestRoute{} to \ResBeamFiveTestRoute{}\,\%. Duration comparisons use the same drives. \\

\specialrule{1.2pt}{0pt}{0pt}

\tint{cPub}
\rqtag{cPub}{RQ3}
& \rqname{cPub}{failure and robustness}
& Methods fail on different windows; mild quantisation leaves the tested attacks effective.
& Per-window choice among ten methods: \CompDTestOracleRoutePct{}\,\% route recovery versus \CompDTestRefRoutePct{}\,\% for the beam. Steering quantised to 10$^\circ$: at most \RobMapSteerMildMaxRouteAbs{} percentage points of change. \\

\arrayrulecolor{TabRule}
\bottomrule
\end{tabularx}

\arrayrulecolor{black}
\end{table*}

Table~\ref{tab:findings} summarises the findings. Table~\ref{tab:main} and
Fig.~\ref{fig:cdf} show the main rows on the five-minute windows of the test drivers (\ResPfFiveTestN{} windows,
\ResPfFiveTestDrivers{} drivers) and few-shot drivers (\ResDrFiveFewShotN{} windows, \ResDrFiveFewShotDrivers{} drivers;
\ResPfFiveFewShotN{} under the map protocols after the map-start rule); all rows but the two area adaptations know the
start position and, unless marked, the heading.

\textbf{Without a map, every CAN-only method drifts far within five minutes.} Nominal steering dead reckoning has
a median ADE of \ResDrFiveTestMed{}\,m on test and \ResDrFiveFewShotMed{}\,m on few-shot drivers, no better
than a straight line at the measured speed (\ResStraightFiveTestMed{}\,m); training-fold calibration does not help on
test (\ResDrCalFiveTestMed{}\,m). The learned GRU has the lowest no-map error (\ResGruFiveTestMed{}\,m on test,
\ResGruFiveFewShotMed{}\,m on few-shot drivers), clearly better than dead reckoning only on the few-shot drivers (on
test the paired CI includes zero: \PairGruVsDrFiveTestMed{}\,m, [\PairGruVsDrFiveTestLo{}, \PairGruVsDrFiveTestHi{}];
Section~\ref{sec:signals}).

\begin{figure*}[htbp]\centering
\includegraphics[width=0.9\textwidth]{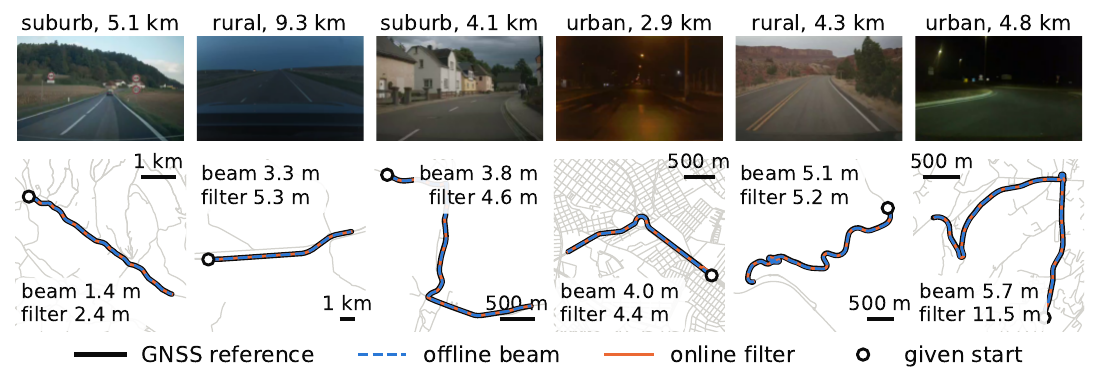}
\caption{Six evaluation five-minute windows recovered by the offline beam and the online filter (speed, steering, start
and heading known) below their camera frames at mid-window (not an input); grey: roads; numbers: ADE. Chosen among
windows both methods recover, across densities and turn counts: accurate, not typical cases (medians in Table~\ref{tab:main}).}\label{fig:examples}
\end{figure*}

\textbf{With the road map, errors drop by more than an order of magnitude.} The online particle filter reaches
\ResPfFiveTestMed{}\,m [\ResPfFiveTestLo{}, \ResPfFiveTestHi{}] and the offline beam \ResBeamFiveTestMed{}\,m
[\ResBeamFiveTestLo{}, \ResBeamFiveTestHi{}], recovering the route in \ResPfFiveTestRoute{}\,\% and
\ResBeamFiveTestRoute{}\,\% of the windows. On the same episodes the filter is closer than dead reckoning in
\PairMapGainFiveTestWinA{}\,\% of the cases (paired median difference \PairMapGainFiveTestMed{}\,m). The driver-equal
medians (\ResPfFiveTestDrv{} and \ResBeamFiveTestDrv{}\,m) agree with the pooled ones at this horizon.
Fig.~\ref{fig:examples} shows six recovered windows.

\textbf{Published methods with the specified priors.} OPD-II reaches \ResOpdTwoFiveTestMed{}\,m on test and
\ResOpdTwoFiveFewShotMed{}\,m on few-shot drivers. Its paired median difference from the map filter includes zero
(\PairOpdVsPfFiveTestMed{}\,m [\PairOpdVsPfFiveTestLo{}, \PairOpdVsPfFiveTestHi{}]), but its paired mean error is
higher by \PairOpdVsPfFiveTestMean{}\,m [\PairOpdVsPfFiveTestMeanLo{}, \PairOpdVsPfFiveTestMeanHi{}] on test.
It recovers \ResOpdTwoFiveTestRoute{}\,\% of test routes versus \ResPfFiveTestRoute{}\,\% for the filter and has more
large endpoint errors. Its strictly online output has nearly the same median
(\ResOpdTwoOnFiveTestMed{}\,m). At fifteen minutes, paired median differences remain inconclusive on both groups;
OPD-II has a higher paired mean error on few-shot drivers and more terminal excursions on both groups.
MacroTracking reaches \ResMacroFiveTestMed{}\,m on test and \ResMacroFiveFewShotMed{}\,m on few-shot drivers; steering
calibration leaves its paired median essentially unchanged (\PairMacroCalVsMacroFiveTestMed{}\,m).

Elastic Pathing uses speed and a known start, without the heading. It reaches \ResEpFiveTestMed{}\,m on test, with
\ResEpFiveTestFail{} failures among \ResEpFiveTestN{} windows. Its median across drivers is substantially higher
(\ResEpFiveTestDrv{}\,m), showing that accurate windows concentrate in some drivers. Giving it the heading reduces
the pooled median to \ResEpHpFiveTestMed{}\,m but increases failures to \ResEpHpFiveTestFail{}: restricting the start
edges removes some successful searches. Dewri et al. reach \ResDewriHpFiveTestMed{}\,m with the heading and
\ResDewriFiveTestMed{}\,m without it. Without the heading, their search often returns the same path as Elastic
Pathing; the paired median difference is \PairDewriVsEpFiveTestMed{}\,m, while the paired mean is higher by
\PairDewriVsEpFiveTestMean{}\,m [\PairDewriVsEpFiveTestMeanLo{}, \PairDewriVsEpFiveTestMeanHi{}].

CAN-Trace finds only \ResCtFiveTestOut{} matches among \ResCtFiveTestN{} pedal-valid test windows with the start
known (few-shot: \ResCtFiveFewShotOut{} of \ResCtFiveFewShotN{}). Its 5\,km area adaptation returns
\ResCtAreaFiveTestOut{} test matches, but the reference leaves that square in \AreaCtTestOutPct{}\,\% of windows,
where recovery is structurally impossible. A second limitation concerns event extraction: unmatched windows contain
only \CtDiagTraceNodesMed{} detected nodes at the median, against about \CtDiagIntersectionsMed{} reference
intersections. These results evaluate our event detector and area adaptation, not CAN-Trace's original
containing-area task.

SPy recovers \ResSpyHpFiveTestRoute{}\,\% of test routes with start and heading and \ResSpyFiveTestRoute{}\,\% with
the start alone. The area variant recovers almost none; \AreaSpyTestOutPct{}\,\% of routes leave its square and
\ResSpyAreaFiveTestFail{} searches exhaust their budget. Even on fifteen-minute windows with at most one reference
turn, the start-and-heading variant has high path precision (\DiagLowTurnPrecSpyHp{}) but low recall
(\DiagLowTurnRecallSpyHp{}): it follows part of the route and stops short. The cause is unresolved. SPy's original
41/58 successful trips used one city's road database and manually trimmed trip ends~\cite{pese2020spy}; differences
in inputs, candidate roads and scoring prevent attributing this performance gap to any one factor.

\textbf{Additional check on controlled drives.} Nine of the methods above (steering dead reckoning; the filter and the beam with
steering and with speed alone; MacroTracking; OPD-II; Elastic Pathing and Dewri et al.\ from the known start and heading)
then ran once, unchanged, on the corpus's controlled supplement: \SuppN{} five-minute windows of \SuppDrives{} drives by
\SuppDrivers{} other drivers in one vehicle and one metropolitan area, never part of the comparison or of any selection. The
order changes: MacroTracking has the lowest median (\SuppMacroMed{}\,m [\SuppMacroLo{}, \SuppMacroHi{}]) and recovers
\SuppMacroRoute{}\,\% of the routes; OPD-II, the beam and the filter reach \SuppOpdTwoMed{}, \SuppBeamMed{} and
\SuppPfMed{}\,m and \SuppOpdTwoRoute{}, \SuppBeamRoute{} and \SuppPfRoute{}\,\%, with wide intervals that overlap one
another (filter [\SuppPfLo{}, \SuppPfHi{}]\,m); steering dead reckoning without a map reaches \SuppDrMed{}\,m
(\ResDrFiveTestMed{}\,m on test) and the speed-only methods \SuppEpHpMed{} (Elastic Pathing) to \SuppBeamNoneMed{}\,m
(beam). With \SuppDrivers{} drivers, one vehicle and one area this cautions against generalising the ranking among the
steering map methods and says nothing about any vehicle or region.

\subsubsection{Contribution of each signal}\label{sec:signals}
\textbf{Does the model use an added signal at all?} On validation chunks of 300\,s we shuffle one input block across
chunks and recompute the fixed learner's loss. Shuffling steering multiplies the loss by \SigUseSteerRatioMin{} to
\SigUseSteerRatioMax{} (all seeds, with or without pedals); shuffling the continuous pedal changes it by
\SigUsePedalSpeedPedalPctMin{} to \SigUsePedalSpeedPedalPctMax{}\,\% next to speed alone and by
\SigUsePedalAllPedalPctMin{} to \SigUsePedalAllPedalPctMax{}\,\% next to speed and steering (three seeds each), the
pressed flags by \SigUseFlagsPctMin{} to \SigUseFlagsPctMax{}\,\% (one seed each): the pedal's effect is small and, next to speed alone, not consistent across seeds.

\textbf{Steering provides the main signal gain.} The controlled study uses \SigSetMapSOneN{} five-minute windows
from \SigNDrivers{} evaluation drivers with a continuous accelerator signal. The learner, decoder and drives are
fixed across signal sets, with three seeds for continuous inputs and one for pressed flags. Adding steering to speed
changes the paired median error by \SigGainMapSteerSpeedMax{} to \SigGainMapSteerSpeedMin{}\,m; every seed's CI
excludes zero, and the median falls from \SigSetMapSOneMed{} to \SigSetMapSThreeMed{}\,m. Adding the continuous
pedal changes it by \SigGainMapPedSpeedMin{} to \SigGainMapPedSpeedMax{}\,m with speed alone (all CIs include zero)
and by \SigGainMapPedAllMin{} to \SigGainMapPedAllMax{}\,m with steering. Pressed flags slightly worsen the
speed-only result (\SigGainMapFlagsSpeedMed{}\,m [\SigGainMapFlagsSpeedLo{}, \SigGainMapFlagsSpeedHi{}]). This subset
is harder for speed-based search than the full evaluation groups, so its gains should be interpreted within the
paired study.

The untrained map filter gives a consistent result on all test windows: \ResPfFiveTestMed{}\,m with steering
versus \ResPfNoneFiveTestMed{}\,m with speed alone (paired difference \PairSteerPfFiveTestMed{}\,m). Without the
heading, these become \ResPfHmFiveTestMed{} and \ResPfNoneHmFiveTestMed{}\,m. Elastic Pathing is substantially better
than the speed-only beam without the heading: paired difference \PairEpVsBeamNoneHmFiveTestMed{}\,m
[\PairEpVsBeamNoneHmFiveTestLo{}, \PairEpVsBeamNoneHmFiveTestHi{}]. Its speed-based turn constraints provide useful
information that the baselines' uniform junction choices omit.

\textbf{A better motion model does not help with the map; without it, the gain is clear only on the few-shot drivers.}
Without the map the learned GRU and the learned
physics residual lower the median error of nominal steering dead reckoning from \ResDrFiveFewShotMed{} to
\ResGruFiveFewShotMed{} and \ResResidFiveFewShotMed{}\,m on few-shot drivers (paired \PairGruVsDrFiveFewShotMed{} and
\PairResidVsDrFiveFewShotMed{}\,m over three seeds, SD \PairGruVsDrFiveFewShotSeedSd{} and
\PairResidVsDrFiveFewShotSeedSd{}\,m; CI excluding zero for \PairGruVsDrFiveFewShotSeedsSig{} and
\PairResidVsDrFiveFewShotSeedsSig{} of \PairGruVsDrFiveFewShotSeeds{} seeds; test CIs include zero). Used as the
motion source of the same map filter, both are slightly worse than nominal steering (few-shot paired medians
\PairGruPfVsPfFiveFewShotMed{} and \PairResidPfVsPfFiveFewShotMed{}\,m, CIs excluding zero for
\PairGruPfVsPfFiveFewShotSeedsSig{} and \PairResidPfVsPfFiveFewShotSeedsSig{} of \PairGruPfVsPfFiveFewShotSeeds{} seeds;
on test \PairGruPfVsPfFiveTestMed{} and \PairResidPfVsPfFiveTestMed{}\,m with CIs including zero). Even the
localizer's GPS-informed curvature, far more accurate without the map (\ResCurvFiveTestMed{} against
\ResDrFiveTestMed{}\,m), gives the filter \ResPfCurvFiveTestMed{}\,m, no better than nominal steering: once the road
network constrains the path, the heading rate's accuracy matters little at five minutes.
\textbf{A CAN yaw rate or a gyroscope gives no reliable gain over steering.} Without the map, the CAN yaw rate, which
only some brands broadcast, differs from steering in opposite directions on the two folds (test, \PairYawVsSteerFiveTestN{}
windows of \PairYawVsSteerFiveTestDrivers{} drivers: \PairYawVsSteerFiveTestMed{}\,m [\PairYawVsSteerFiveTestLo{},
\PairYawVsSteerFiveTestHi{}]; few-shot, \PairYawVsSteerFiveFewShotN{} of \PairYawVsSteerFiveFewShotDrivers{}:
\PairYawVsSteerFiveFewShotMed{}\,m [\PairYawVsSteerFiveFewShotLo{}, \PairYawVsSteerFiveFewShotHi{}]); with so few
drivers this is inconclusive. The driver-assistance system's gyroscope (not a CAN signal) is indistinguishable from
steering on both folds (paired medians \PairGyroVsSteerFiveTestMed{} and \PairGyroVsSteerFiveFewShotMed{}\,m on
\PairGyroVsSteerFiveTestN{} test and \PairGyroVsSteerFiveFewShotN{} few-shot windows; both CIs include zero).

\subsection{RQ2: Initial heading, offline access and duration}\label{sec:rq2}
\textbf{With steering, the start heading barely moves the offline beam's median, unlike the filter's; it still
matters in the tail and, with speed alone, for the beam too.} Without the start heading the online
filter's five-minute median grows from \ResPfFiveTestMed{} to \ResPfHmFiveTestMed{}\,m on test (paired median
difference \PairHeadPfFiveTestMed{}\,m [\PairHeadPfFiveTestLo{}, \PairHeadPfFiveTestHi{}]), while the offline beam
moves only from \ResBeamFiveTestMed{} to \ResBeamHmFiveTestMed{}\,m (paired median difference \PairHeadBeamFiveTestMed{}\,m;
knowing the heading makes it closer in only \PairHeadBeamFiveTestWinA{}\,\% of the episodes). Where it helps, the gain is large: the paired mean difference is \PairHeadBeamFiveTestMean{}\,m
[\PairHeadBeamFiveTestMeanLo{}, \PairHeadBeamFiveTestMeanHi{}] (filter \PairHeadPfFiveTestMean{}\,m), and the share of
routes recovered falls from \ResBeamFiveTestRoute{} to \ResBeamHmFiveTestRoute{}\,\% (filter \ResPfFiveTestRoute{} to
\ResPfHmFiveTestRoute{}\,\%).
With speed alone the offline beam depends strongly on the heading (\ResBeamNoneFiveTestMed{} against
\ResBeamNoneHmFiveTestMed{}\,m). The same holds on fifteen-minute windows (\ResBeamFifteenTestMed{} vs \ResBeamHmFifteenTestMed{}\,m; paired means in the released tables). A decoder that sees the whole window can test every start direction against the later turns; the online
filter, plausibly, must spread its particles over them until the first turn tells them apart.
\textbf{With the heading known, offline access buys little at the median, about half of it through the output.} On
test windows the paired median difference between the offline beam and the online filter is \PairBeamVsPfFiveTestMed{}\,m
[\PairBeamVsPfFiveTestLo{}, \PairBeamVsPfFiveTestHi{}]. The heaviest-lineage output of the same filter (same
particles and motion model) differs from the filter by a paired median of \PairLineageVsPfFiveTestMed{}\,m
[\PairLineageVsPfFiveTestLo{}, \PairLineageVsPfFiveTestHi{}]; beam and lineage differ by \PairBeamVsLineageFiveTestMed{}\,m
[\PairBeamVsLineageFiveTestLo{}, \PairBeamVsLineageFiveTestHi{}] (CI includes zero) at the median and in the tail on
fifteen-minute few-shot windows (paired mean \PairBeamVsLineageFifteenFewShotMean{}\,m
[\PairBeamVsLineageFifteenFewShotMeanLo{}, \PairBeamVsLineageFifteenFewShotMeanHi{}]; test CI includes zero).

\begin{figure*}[!tp]\centering
\includegraphics[width=0.8\textwidth]{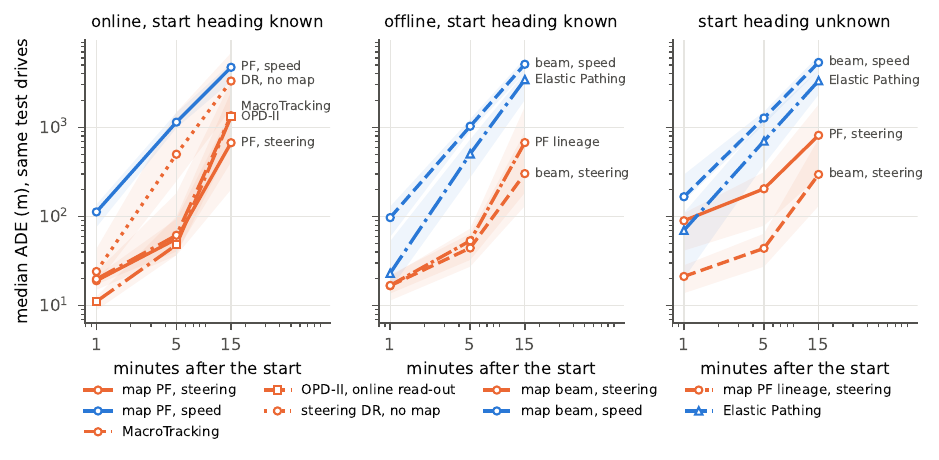}
\caption{The same \CohTestN{} test drives (\CohTestDrivers{} drivers), one start each, scored 1, 5 and 15 minutes after
the start: median ADE with driver-resampling 95\,\% bands. Online methods are scored on prefixes of their 15-minute
runs; offline methods are rerun on each prefix with only its signals and map. Encoding as in Fig.~\ref{fig:cdf}.}\label{fig:horizon}
\end{figure*}

\textbf{Horizon on one cohort.} The cohort is the first fifteen-minute window of each of \NCohortDrives{} test and
few-shot drives (\NCohortDrivers{} drivers), cut at 1, 5 and 15 minutes from the same start (Fig.~\ref{fig:horizon}). On the \CohTestN{} test drives,
OPD-II has the lowest median after one minute (\CohOpdTwoOnlineOMHpOneTestMed{}\,m, unpaired; the map filter
\CohMapPfSteerOMHpOneTestMed{}\,m), but the median of the offline beam with steering grows least: \CohMapBeamSteerFMHpOneTestMed{},
\CohMapBeamSteerFMHpFiveTestMed{} and \CohMapBeamSteerFMHpFifteenTestMed{}\,m at 1, 5 and 15 minutes, against
\CohMapPfSteerOMHpOneTestMed{}, \CohMapPfSteerOMHpFiveTestMed{} and \CohMapPfSteerOMHpFifteenTestMed{}\,m for the online
filter and \CohOpdTwoFMHpFifteenTestMed{} and \CohMacrotrackingOMHpFifteenTestMed{}\,m for OPD-II and MacroTracking at
fifteen minutes (unpaired). Paired on the same windows, beam and filter agree on most of
them (paired median difference at fifteen minutes \CohPairBeamVsPfFifteenTestMed{}\,m [\CohPairBeamVsPfFifteenTestLo{},
\CohPairBeamVsPfFifteenTestHi{}]) and the beam avoids the filter's largest failures (mean difference
\CohPairBeamVsPfFifteenTestMean{}\,m [\CohPairBeamVsPfFifteenTestMeanLo{}, \CohPairBeamVsPfFifteenTestMeanHi{}]); the
lineage output behaves like the filter here (paired mean against the beam \CohPairBeamVsLineageFifteenTestMean{}\,m
[\CohPairBeamVsLineageFifteenTestMeanLo{}, \CohPairBeamVsLineageFifteenTestMeanHi{}]), so the beam's search, not its single-path output, avoids these failures. The speed-only methods are kilometres off after fifteen minutes (Elastic Pathing \CohElasticPathingFMHmFifteenTestMed{}\,m,
\CohElasticPathingFMHmFifteenTestFail{} of the \CohTestN{} drives without a complete path; beam
\CohMapBeamNoneFMHmFifteenTestMed{}\,m). This cohort is harder than the pooled fifteen-minute comparison (filter \ResPfFifteenTestMed{}\,m there).

\textbf{Whole drives.} Drive duration differs between the two evaluation groups (median \FullDurMedTest{} versus
\FullDurMedFewShot{} minutes), so results are reported separately. In \FullBeyondDiscTest{} of \FullMapOkTest{} test
and \FullBeyondDiscFewShot{} of \FullMapOkFewShot{} few-shot drives, the endpoint lies outside the 60\,km map disc.
Methods restricted to that disc cannot recover the tail. The steering beam has the lowest CAN-only median on both
groups (\ResBeamFullTestMed{} and \ResBeamFullFewShotMed{}\,m); removing the heading gives
\ResBeamHmFullTestMed{} and \ResBeamHmFullFewShotMed{}\,m. The ranking of the filter, MacroTracking and OPD-II differs
between groups and is descriptive: intervals overlap, these comparisons are unpaired, and Meili's map access differs.
Speed-only searches are kilometres off, while SPy with start and heading recovers only
\ResSpyHpFullTestRoute{} and \ResSpyHpFullFewShotRoute{}\,\% of whole-drive routes.

\subsection{RQ3: Failures, cost and robustness}\label{sec:rq3}
\textbf{Without steering, a good candidate is often present but ranked low, and more often absent.} The offline beam
keeps ten complete paths (the best-of-$K$ curves are in the released figure data).
With speed alone and no start heading its top-ranked path has a median error of \TopkMapBeamNoneHmTestKOne{}\,m on
test, the best of its top three \TopkMapBeamNoneHmTestKThree{}\,m and the best of ten
\TopkMapBeamNoneHmTestKTen{}\,m; the best path is ranked first in only \TopkMapBeamNoneHmTestBestFirst{}\,\% of the
windows. With steering the gap closes (\TopkMapBeamSteerHmTestKOne{}\,m at rank one, \TopkMapBeamSteerHmTestKTen{}\,m
for the best of ten, best path first in \TopkMapBeamSteerHmTestBestFirst{}\,\%). Elastic Pathing's
stopping rule usually leaves a single complete path (\TopkElasticPathingHmTestSingle{} of its
\TopkElasticPathingHmTestOut{} test outputs), so its top-$K$ adds little (\TopkElasticPathingHmTestKOne{} to
\TopkElasticPathingHmTestKTen{}\,m). Dewri et al.'s ten candidates behave like the speed-only beam's (\TopkDewriHmTestKOne{} to
\TopkDewriHmTestKTen{}\,m, best first in \TopkDewriHmTestBestFirst{}\,\%).
Splitting the test windows by a 50\,m whole-path criterion separates the two failure kinds: with speed alone and no
start heading, the beam's top-ranked path is within 50\,m of the reference in \TaxMapBeamNoneHmTestFiftyTopOneOkPct{}\,\%
of the windows, a lower-ranked candidate is while the top one is not in \TaxMapBeamNoneHmTestFiftyRankMissPct{}\,\%
(a ranking failure), and no candidate is in \TaxMapBeamNoneHmTestFiftyNoCandPct{}\,\% (a search failure); with
steering the shares are \TaxMapBeamSteerHpTestFiftyTopOneOkPct{}, \TaxMapBeamSteerHpTestFiftyRankMissPct{} and
\TaxMapBeamSteerHpTestFiftyNoCandPct{}\,\%. The top-$K$ gains come from the ranking
failures; most speed-only windows have no good candidate at all.

\textbf{The beam's candidate score does not tell an attacker when it is right, unless speed is all it has.} Each beam
candidate carries a score normalised over the surviving candidates (a confidence score, not a calibrated probability):
with one path or near-duplicates the score is 1 whatever the outcome (\ConfBeamHpFiveTestProbOnePct{}\,\% of the test
windows with steering). With steering the Top-1 score separates
recovered from unrecovered routes no better than chance (AUROC \ConfBeamHpFiveTestAuroc{}, CI
[\ConfBeamHpFiveTestAurocLo{}, \ConfBeamHpFiveTestAurocHi{}] on test, \ConfBeamHpFiveFewShotAuroc{} on few-shot drivers;
Spearman correlation with the ADE \ConfBeamHpFiveTestRho{}): a threshold of 0.9 accepts \ConfBeamHpFiveTestNinetyCovPct{}\,\% of the test windows at a route rate of
\ConfBeamHpFiveTestNinetyRoutePct{}\,\%, and
\ConfBeamHpFiveTestNinetyErrPct{}\,\% of the accepted routes are wrong. With speed alone the candidates differ and the
score is informative (AUROC \ConfBeamNoneHpFiveTestAuroc{} with the heading, \ConfBeamNoneHmFiveTestAuroc{} without;
a threshold of 0.9 accepts \ConfBeamNoneHpFiveTestNinetyCovPct{}\,\% of the windows at a route rate of
\ConfBeamNoneHpFiveTestNinetyRoutePct{}\,\% against \ResBeamNoneFiveTestRoute{}\,\% overall). Elastic Pathing and
Dewri et al.\ give a score of 1 to \ConfProbOneEpHpPct{} and \ConfProbOneDewriHpPct{}\,\% of their outputs (one path, or
uniformly scored paths), so no selection is possible there.

\textbf{With steering, the filter's and the beam's losses start disproportionately just after turns; the speed-only
filter's do not.} A fifteen-minute window is lost when it ends in a terminal excursion (Section~\ref{sec:problem}). With steering, \FailPfLostPct{}\,\% of the filter's \FailPfN{} fifteen-minute windows (test and few-shot
pooled) and \FailBeamLostPct{}\,\% of the beam's end lost, and the onsets cluster just after turns: \FailPfAfterTurnPct{}\,\% and
\FailBeamAfterTurnPct{}\,\% of them fall within 30\,s after a reference turn, against about \FailPfChancePct{} and
\FailBeamChancePct{}\,\% if onsets were spread evenly over the window; this is an association (most onsets fall
elsewhere; folds pooled; the published methods and the speed-only beam were not analysed). The speed-only filter is lost in
\FailPfNoneLostPct{}\,\% of the windows, with no such link to turns (\FailPfNoneAfterTurnPct{} against
\FailPfNoneChancePct{}\,\%).

\textbf{The speed-only map filter has much larger errors in denser road networks; the steering methods differ less,
and mainly over longer horizons.} Split by road density around the start (tertiles at \DensCutLow{} and \DensCutHigh{}
nodes per km$^2$; folds pooled; \DensNFive{} windows of \DensMapPfSteerLowDrivers{}, \DensMapPfSteerMidDrivers{} and
\DensMapPfSteerHighDrivers{} drivers per tertile, sparse to dense), the filter with
steering has five-minute medians of \DensMapPfSteerLowMed{}, \DensMapPfSteerMidMed{} and \DensMapPfSteerHighMed{}\,m from
sparse to dense, while the speed-only filter goes from \DensMapPfNoneLowMed{} to \DensMapPfNoneMidMed{} and
\DensMapPfNoneHighMed{}\,m. The steering filter is not unaffected: its five-minute tail-lost share rises from
\DensMapPfSteerLowLostPct{} to \DensMapPfSteerHighLostPct{}\,\% and its fifteen-minute median (\DensNFifteen{} windows per
tertile) from \DensMapPfSteerLowMedFifteen{} to \DensMapPfSteerHighMedFifteen{}\,m (beam \DensMapBeamSteerLowMedFifteen{}
to \DensMapBeamSteerHighMedFifteen{}\,m). Density is
confounded with region, driver, speed and fold, so these are associations, not effects.

\textbf{The methods fail on different windows; almost no window defeats all of them, and the common failures sit in
the windows with many turns.} For each five-minute window we take the best of the ten map methods with the start
heading, by the lowest ADE or by route recovery (two oracles).
Some method recovers the route in \CompDTestOracleRoutePct{}\,\% of the test windows (few-shot
\CompDFewShotOracleRoutePct{}\,\%) against \CompDTestRefRoutePct{}\,\% for the beam, our fixed reference, a difference of
\CompDTestRouteGainPct{} points (CI [\CompDTestRouteGainLo{}, \CompDTestRouteGainHi{}]); none of them recovers it in
\CompDTestAllFailPct{}\,\% of the windows (CI [\CompDTestAllFailLo{}, \CompDTestAllFailHi{}]). This common failure grows
from \CFailTurnsZeroOnePct{}\,\% of the windows with at most one reference turn to \CFailTurnsSevenPlusPct{}\,\% with seven
or more (test and few-shot pooled, \CFailTurnsSevenPlusN{} windows) and from \CFailDensLowPct{} to \CFailDensHighPct{}\,\%
across the density tertiles. No method attains the per-window minimum ADE in more than \CompMemDTestOpdTwoAttainPct{}\,\% of
the windows (OPD-II; lineage \CompMemDTestLineageAttainPct{}, beam \CompMemDTestBeamAttainPct{}\,\%); the oracle improves
on the beam's ADE by a paired median of \CompDTestAdeGainMed{}\,m [\CompDTestAdeGainLo{}, \CompDTestAdeGainHi{}]. These are facts
about the current method set, not about what is recoverable: methods that share a motion model, a map or a search
share failures too. Per driver (\DrvRouteBeamTestNDriversGeFive{} test drivers with at least five windows), the beam's route share has a
median of \DrvRouteBeamTestMed{}\,\% and is below 50\,\% for \DrvRouteBeamTestBelowHalfPct{}\,\% of them.

\subsubsection{Computational cost and degraded signals}\label{sec:cost}
\textbf{Inference is usually inexpensive; map preparation is a separate cost.} Timing uses a fixed validation
subset of \CostN{} windows, with one method and one worker at a time; failed searches retain their elapsed time.
For five-minute windows, median times are \CostDrSteerFive{}\,s for steering dead reckoning,
\CostMacrotrackingFive{}\,s for MacroTracking, \CostOpdTwoFive{}\,s for OPD-II, \CostMapBeamSteerFive{}\,s for the
beam and \CostMapPfSteerFive{}\,s for the filter. Elastic Pathing has a long tail (90th percentile
\CostElasticPathingFiveNinety{}\,s). SPy's area search takes \CostSpyAreaFive{}\,s at the median and exhausts its
budget on \CostSpyAreaFailFive{} of \CostNFive{} windows. CAN-Trace is timed separately on
\CostPedalNFive{} pedal-valid windows: \CostPedalCanTraceFive{}\,s with the start known and
\CostPedalCanTraceAreaFive{}\,s with the area prior, usually without a match. Building a drive's graph takes
\CostGraph{}\,s at the median; peak memory is at most \CostRssMax{}\,GB. Preparing all road tiles takes
\PrepTilesMin{} minutes, while Valhalla preparation takes \PrepValhallaMin{} minutes and
\PrepValhallaGb{}\,GB on disk. Smaller particle and beam budgets lose little accuracy in the supplementary comparisons.

\textbf{Degraded signals barely change dead reckoning, and the fixed map attacks tolerate most coarsening.} Quantising
the speed to 1\,m/s, adding 0.5\,m/s of speed noise or sampling the steering at 1\,Hz moves the paired median error of
steering dead reckoning by at most \RobustDrMaxAbs{}\,m on either fold, against errors of hundreds of metres. For the
map attacks we re-ran the steering filter and beam and the speed-only beam on the five-minute windows of both evaluation
folds with one signal degraded before the method sees it, every parameter unchanged, on a grid fixed before the first
run (per-level curves in the released figure data): steering-wheel angle quantised to 2, 5, 10 or 20$^\circ$, held at 1 or
0.2\,Hz or offset by a per-second noise of 2 or 5$^\circ$ (one seed); speed quantised to 1, 2 or 5\,m/s or held at
1\,Hz. Steering quantised to 10$^\circ$, offset by 5$^\circ$ or held at 1\,Hz changes the
route recovery of either attack by at most \RobMapSteerMildMaxRouteAbs{} percentage points; at 20$^\circ$ both lose about one point
(filter \RobMapPfSqTwentyTestRouteDiff{}
[\RobMapPfSqTwentyTestRouteDiffLo{}, \RobMapPfSqTwentyTestRouteDiffHi{}] on test). Steering held at 0.2\,Hz changes the filter's
route recovery by \RobMapPfSFifthHzTestRouteDiff{} and \RobMapPfSFifthHzFewShotRouteDiff{} points (test, few-shot), the
beam's by \RobMapBeamSFifthHzFewShotRouteDiff{} on few-shot, and raises the filter's test median from
\RobMapPfSFifthHzTestMedClean{} to \RobMapPfSFifthHzTestMed{}\,m. Speed quantised to 1\,m/s changes no route rate measurably (within
\RobMapPqOneMaxRouteAbs{} points, every CI including zero); at 5\,m/s (18\,km/h) the
two steering attacks change their route recovery by \RobMapBeamPqFiveTestRouteDiff{} (beam, test) to
\RobMapPfPqFiveFewShotRouteDiff{} (filter, few-shot) points and their medians double or triple, whereas the speed-only
beam's route rate does not change (\RobMapBeamNonePqFiveTestRouteDiff{} [\RobMapBeamNonePqFiveTestRouteDiffLo{},
\RobMapBeamNonePqFiveTestRouteDiffHi{}] on test). Even at the harshest levels the attacks recover
\RobMapHardMinRoute{}\,\% or more of the routes. Delaying the steering by 0.5\,s (a timing probe, reported apart) does
not degrade either attack (route recovery rises by about one point, as with steering held at 1\,Hz: the vehicle's yaw
lags the steering-wheel angle, which the nominal bicycle model ignores). These are
sensitivities of fixed attacks, none re-tuned, re-calibrated or re-trained on the degraded signals: they bound what a
publisher gains from coarsening against these attacks, not against an adapted one, and set no protection threshold.

\input{tables/agent.tex}
\subsubsection{LLM-based candidate selection}\label{sec:agent}
We test whether a language model can select better reconstructions from existing tools. The study uses the first
eligible five-minute window from each evaluation drive: \AgentCohortTestN{} test and
\AgentCohortFewShotN{} few-shot windows, with speed, steering, map, start and heading. The baseline is the beam
alone (A). A supervised logistic selector (B) and Qwen3-14B agent (C) choose among the beam, the filter's best lineage
and Elastic Pathing. The selector uses validation labels; the agent uses tool diagnostics and can return only a
candidate it has seen. Model settings and budgets are fixed after a validation pilot (three seeds; four turns and
seven tool calls). An extended agent (C$^{+}$) has ten reconstructor presets, comparison and scoring tools, twelve
turns and forty calls; B$^{+}$ runs and scores every preset. Full prompts and settings are in the supplement.

On test drivers, route recovery is \AgentBeamTestRoute{}\,\% for the beam, \AgentSelTestRoute{}\,\% for the selector
and \AgentLlmTestRoute{}\,\% for the base agent (Table~\ref{tab:agent}). The agent-minus-selector difference is
\AgentPairLlmVsSelTestRouteDiff{} percentage points [\AgentPairLlmVsSelTestRouteLo{},
\AgentPairLlmVsSelTestRouteHi{}] on test and \AgentPairLlmVsSelFewShotRouteDiff{}
[\AgentPairLlmVsSelFewShotRouteLo{}, \AgentPairLlmVsSelFewShotRouteHi{}] on few-shot drivers. Thus the base agent has
no clear advantage over the selector, while its estimated serial cost is \AgentLlmTestCost{}\,s per window versus
\AgentSelTestCost{}\,s. The extended agent reaches \AgentMaxLlmTestRoute{}\,\% on test but
\AgentMaxLlmFewShotRoute{}\,\% on few-shot drivers; its combined recovery does not improve over the base agent
(\AgentMaxPairLlmVsPrevAllRouteDiff{} points [\AgentMaxPairLlmVsPrevAllRouteLo{},
\AgentMaxPairLlmVsPrevAllRouteHi{}]). On test drivers, the extended agent improves over its selector by \AgentMaxPairLlmVsSelTestRouteDiff{} points [\AgentMaxPairLlmVsSelTestRouteLo{}, \AgentMaxPairLlmVsSelTestRouteHi{}], but the improvement does not extend to the few-shot group. An oracle choosing from the available candidates would recover
\AgentOracleTestRoute{}\,\% of test routes with three tools and \AgentMaxOracleTestRoute{}\,\% with all presets,
indicating substantial room to improve selection. These results concern one model and two configurations.

%% file: tables/main_board.tex
\begin{table*}[!tp]
\centering

\definecolor{cHeat}{HTML}{2BB673}
\providecommand{\ci}[1]{\textcolor{black!60}{#1}}

\makeatletter
\providecommand{\hc}[1]{%
  \cellcolor{cHeat!\strip@pt\dimexpr #1pt*28\relax}#1%
}
\makeatother

\caption{Five-minute results. ADE: median average displacement error (m) with 95\,\% driver-bootstrap CI; $c_{100}$: fraction of seconds within 100\,m; route: fraction with path recall@50\,m $\ge0.8$. Missing outputs count as infinite ADE and zero coverage or recovery. The start and heading are given unless a method or group is marked otherwise. Learned results average three seeds. Driver-level medians, seed variation and failure counts are in the supplementary tables.}
\label{tab:main}

\small
\setlength{\tabcolsep}{5.5pt}
\renewcommand{\arraystretch}{1.0}
\setlength{\aboverulesep}{0pt}
\setlength{\belowrulesep}{0pt}

\begin{threeparttable}
\begin{tabularx}{\textwidth}{
  >{\raggedright\arraybackslash}X
  r@{\hspace{6pt}}l
  c c
  c
  r@{\hspace{6pt}}l
  c c
}

\rowcolor{cHead}
 & \multicolumn{4}{l}{\hdspan{Test drivers}}
 &
 & \multicolumn{4}{l}{\hdspan{Few-shot drivers}} \\

\rowcolor{cHead}
\hd{Method}
& \multicolumn{2}{c}{\hd{ADE [CI]}}
& \hd{$\boldsymbol{c_{100}}$}
& \hd{Route}
&
& \multicolumn{2}{c}{\hd{ADE [CI]}}
& \hd{$\boldsymbol{c_{100}}$}
& \hd{Route} \\

\grp{cBase}{No map, online (O-N-H$^+$)}

Hold start
& 3,314 & \ci{[2,113, 4,313]} & \hc{0.03} & \hc{0.00}
& & 2,709 & \ci{[2,232, 3,562]} & \hc{0.03} & \hc{0.00} \\

\tint{cBase}
Straight line (speed)
& 735 & \ci{[644, 911]} & \hc{0.30} & \hc{0.09}
& & 825 & \ci{[781, 951]} & \hc{0.28} & \hc{0.07} \\

Start prior
& 1,068 & \ci{[920, 1,119]} & \hc{0.12} & \hc{0.02}
& & 1,062 & \ci{[946, 1,162]} & \hc{0.13} & \hc{0.02} \\

\tint{cBase}
Route retrieval
& 3,083 & \ci{[2,002, 4,296]} & \hc{0.08} & \hc{0.04}
& & 2,530 & \ci{[2,016, 3,258]} & \hc{0.08} & \hc{0.05} \\

Steering DR, nominal
& 921 & \ci{[461, 1,742]} & \hc{0.25} & \hc{0.04}
& & 1,669 & \ci{[906, 2,472]} & \hc{0.20} & \hc{0.02} \\

\tint{cBase}
Steering DR, online offset
& 869 & \ci{[611, 1,306]} & \hc{0.25} & \hc{0.03}
& & 1,423 & \ci{[964, 1,979]} & \hc{0.20} & \hc{0.02} \\

Steering DR, train calibrated
& 1,312 & \ci{[409, 4,189]} & \hc{0.24} & \hc{0.04}
& & 1,651 & \ci{[919, 2,430]} & \hc{0.20} & \hc{0.02} \\

\tint{cBase}
Learned GRU
& 543 & \ci{[226, 2,478]} & \hc{0.32} & \hc{0.07}
& & 393 & \ci{[305, 492]} & \hc{0.33} & \hc{0.05} \\

Learned physics residual
& 690 & \ci{[310, 2,994]} & \hc{0.29} & \hc{0.06}
& & 435 & \ci{[366, 536]} & \hc{0.32} & \hc{0.04} \\

\grp{cMap}{Map, online (O-M-H$^+$)}

Map PF, speed only
& 286 & \ci{[50, 868]} & \hc{0.55} & \hc{0.56}
& & 618 & \ci{[186, 890]} & \hc{0.52} & \hc{0.48} \\

\tint{cMap}
Map PF, steering
& 43 & \ci{[29, 59]} & \hc{0.83} & \hc{0.87}
& & 27 & \ci{[19, 47]} & \hc{0.88} & \hc{0.90} \\

Map PF, learned GRU
& 89 & \ci{[42, 168]} & \hc{0.65} & \hc{0.86}
& & 42 & \ci{[29, 79]} & \hc{0.79} & \hc{0.86} \\

\tint{cMap}
Map PF, learned residual
& 87 & \ci{[45, 262]} & \hc{0.62} & \hc{0.86}
& & 53 & \ci{[33, 84]} & \hc{0.78} & \hc{0.86} \\

MacroTracking
& 84 & \ci{[66, 210]} & \hc{0.66} & \hc{0.62}
& & 52 & \ci{[40, 143]} & \hc{0.73} & \hc{0.65} \\

\tint{cMap}
MacroTracking, calibrated
& 76 & \ci{[61, 172]} & \hc{0.68} & \hc{0.65}
& & 48 & \ci{[38, 126]} & \hc{0.74} & \hc{0.67} \\

OPD-II, online read-out
& 42 & \ci{[27, 69]} & \hc{0.80} & \hc{0.79}
& & 24 & \ci{[15, 54]} & \hc{0.83} & \hc{0.78} \\

\grp{cLearn}{Map, offline (F-M-H$^+$)}

Map beam, speed only
& 165 & \ci{[37, 633]} & \hc{0.57} & \hc{0.54}
& & 343 & \ci{[88, 637]} & \hc{0.54} & \hc{0.47} \\

\tint{cLearn}
Map beam, steering
& 37 & \ci{[24, 54]} & \hc{0.84} & \hc{0.88}
& & 21 & \ci{[17, 40]} & \hc{0.90} & \hc{0.92} \\

Map PF, best lineage
& 39 & \ci{[28, 56]} & \hc{0.83} & \hc{0.87}
& & 24 & \ci{[17, 44]} & \hc{0.89} & \hc{0.90} \\

\tint{cLearn}
Elastic Pathing, heading given
& 87 & \ci{[38, 324]}$^{\dagger}$ & \hc{0.63} & \hc{0.58}
& & 80 & \ci{[54, 165]}$^{\dagger}$ & \hc{0.66} & \hc{0.59} \\

OPD-II (3\,s delayed)
& 42 & \ci{[27, 68]} & \hc{0.80} & \hc{0.79}
& & 24 & \ci{[15, 54]} & \hc{0.83} & \hc{0.78} \\

\tint{cLearn}
Dewri et al., heading given
& 81 & \ci{[38, 215]}$^{\dagger}$ & \hc{0.65} & \hc{0.62}
& & 89 & \ci{[59, 157]}$^{\dagger}$ & \hc{0.65} & \hc{0.59} \\

SPy, start and heading given
& 1,474 & \ci{[1,283, 1,602]}$^{\dagger}$ & \hc{0.12} & \hc{0.20}
& & 1,284 & \ci{[1,090, 1,596]}$^{\dagger}$ & \hc{0.15} & \hc{0.22} \\

\grp{cPub}{Map, heading unknown (O/F-M-H$^-$)}

Map PF, steering
& 138 & \ci{[99, 175]} & \hc{0.70} & \hc{0.79}
& & 88 & \ci{[61, 124]} & \hc{0.76} & \hc{0.82} \\

\tint{cPub}
Map PF, speed only
& 1,773 & \ci{[1,550, 1,987]} & \hc{0.35} & \hc{0.34}
& & 1,868 & \ci{[1,572, 2,097]} & \hc{0.32} & \hc{0.29} \\

Map beam, steering
& 41 & \ci{[26, 59]} & \hc{0.77} & \hc{0.81}
& & 23 & \ci{[17, 42]} & \hc{0.84} & \hc{0.87} \\

\tint{cPub}
Map beam, speed only
& 1,314 & \ci{[866, 1,619]} & \hc{0.35} & \hc{0.32}
& & 1,527 & \ci{[1,294, 1,738]} & \hc{0.33} & \hc{0.29} \\

Elastic Pathing
& 129 & \ci{[42, 691]}$^{\dagger}$ & \hc{0.56} & \hc{0.53}
& & 177 & \ci{[94, 344]}$^{\dagger}$ & \hc{0.56} & \hc{0.50} \\

\tint{cPub}
CAN-Trace, start known
& \textcolor{cPub}{\textit{fail}}
& \ci{[n/a, n/a]}$^{\dagger}$ & \hc{0.00} & \hc{0.00}
&
& \textcolor{cPub}{\textit{fail}}
& \ci{[n/a, n/a]}$^{\dagger}$ & \hc{0.00} & \hc{0.00} \\

CAN-Trace, 5\,km area
& \textcolor{cPub}{\textit{fail}}
& \ci{[4,344, n/a]}$^{\dagger}$ & \hc{0.00} & \hc{0.00}
&
& \textcolor{cPub}{\textit{fail}}
& \ci{[n/a, n/a]}$^{\dagger}$ & \hc{0.00} & \hc{0.00} \\

\tint{cPub}
Dewri et al.
& 293 & \ci{[81, 775]}$^{\dagger}$ & \hc{0.49} & \hc{0.46}
& & 487 & \ci{[272, 708]}$^{\dagger}$ & \hc{0.48} & \hc{0.43} \\

SPy, start known
& 2,198 & \ci{[1,702, 2,636]}$^{\dagger}$ & \hc{0.08} & \hc{0.11}
& & 1,810 & \ci{[1,512, 2,211]}$^{\dagger}$ & \hc{0.11} & \hc{0.13} \\

\tint{cPub}
SPy, 5\,km area
& 3,929 & \ci{[3,222, 4,406]}$^{\dagger}$ & \hc{0.01} & \hc{0.00}
& & 3,400 & \ci{[3,050, 3,953]}$^{\dagger}$ & \hc{0.01} & \hc{0.00} \\

\grp{cRef}{Reference (GPS-informed class F)}

Curvature DR (class F)
& 84 & \ci{[74, 107]} & \hc{0.64} & \hc{0.27}
& & 80 & \ci{[70, 92]} & \hc{0.67} & \hc{0.27} \\

\tint{cRef}
Map PF, curvature (class F)
& 44 & \ci{[29, 61]} & \hc{0.82} & \hc{0.86}
& & 28 & \ci{[19, 49]} & \hc{0.87} & \hc{0.89} \\

\arrayrulecolor{cHead}
\specialrule{1pt}{0pt}{0pt}
\arrayrulecolor{black}

\end{tabularx}

\begin{tablenotes}[flushleft]
\footnotesize

\item[]
Green shading in $c_{100}$ and Route cells increases with the value.

\item[]
Start $p_0$ and heading $\theta_0$ are given unless noted;
area variants receive a 5\,km square centred on $p_0$.
$^\dagger$~At least one expected window has no output
(counted as $\text{ADE}=\infty$ and zero coverage/recovery).
CAN-Trace uses only pedal-valid drives; OPD-II uses the Meili
map access in Table~\ref{tab:methods}.

\end{tablenotes}
\end{threeparttable}
\end{table*}

%% file: tables/agent.tex
\begin{table*}[!tp]
\centering
\providecommand{\ci}[1]{\textcolor{black!60}{#1}}
\providecommand{\cfg}[2]{%
  {\setlength{\fboxsep}{1.3pt}%
   \colorbox{#1!18}{%
     \makebox[1.7em]{%
       \textcolor{#1!78!black}{%
         \scriptsize\sffamily\bfseries\strut #2}}}}%
}
\providecommand{\costbar}[1]{%
  \textcolor{cOpt!65!white}{%
    \rule[0.05ex]{\dimexpr #1pt*105/256\relax}{1ex}}%
}

\caption{Candidate selection on one five-minute window per evaluation drive across base ($A$--$C$) and extended ($B^+, C^+$) configurations. Columns report route recovery, failure-aware median ADE, and median reconstructed serial cost (95\,\% driver-clustered CIs; three seeds pooled for Qwen3-14B agent $C/C^+$).}
\label{tab:agent}
\footnotesize
\setlength{\tabcolsep}{3.5pt}
\renewcommand{\arraystretch}{1.25}
\setlength{\aboverulesep}{0pt}\setlength{\belowrulesep}{0pt}
\begin{threeparttable}
\begin{tabularx}{0.92\textwidth}{c >{\raggedright\arraybackslash}X r@{\hspace{5pt}}l r@{\hspace{5pt}}l r l c r@{\hspace{5pt}}l r@{\hspace{5pt}}l r l}
\rowcolor{cHead}
 & & \multicolumn{6}{l}{\hdspan{Test drivers ($n=268$)}} & & \multicolumn{6}{l}{\hdspan{Few-shot drivers ($n=391$)}} \\
\rowcolor{cHead}
\multicolumn{2}{l}{\hd{Method}}
 & \multicolumn{2}{c}{\hd{Route \%}} & \multicolumn{2}{c}{\hd{ADE (m)}} & \multicolumn{2}{c}{\hd{Cost (s)}} &
 & \multicolumn{2}{c}{\hd{Route \%}} & \multicolumn{2}{c}{\hd{ADE (m)}} & \multicolumn{2}{c}{\hd{Cost (s)}} \\
\cfg{cBase}{A} & beam & 69 & \ci{[58, 85]} & 49 & \ci{[29, 66]} & 0.3 & \costbar{0.3} & & 83 & \ci{[76, 90]} & 24 & \ci{[18, 36]} & 0.3 & \costbar{0.3} \\
\tint{cMap} \cfg{cMap}{B} & selector & 75 & \ci{[67, 87]} & 38 & \ci{[29, 50]} & 1.1 & \costbar{1.1} & & 83 & \ci{[78, 90]} & 24 & \ci{[18, 34]} & 1.0 & \costbar{1.0} \\
\tint{cLearn} \cfg{cLearn}{C} & agent & 75 & \ci{[66, 88]} & 41 & \ci{[29, 53]} & 36.4 & \costbar{36.4} & & 84 & \ci{[77, 91]} & 23 & \ci{[18, 34]} & 35.0 & \costbar{35.0} \\
\tint{cMap} \cfg{cMap}{B\textsuperscript{+}} & selector, every preset & 74 & \ci{[66, 85]} & 43 & \ci{[28, 59]} & 4.8 & \costbar{4.8} & & 82 & \ci{[75, 90]} & 26 & \ci{[20, 41]} & 4.4 & \costbar{4.4} \\
\tint{cLearn} \cfg{cLearn}{C\textsuperscript{+}} & agent, extended & 78 & \ci{[71, 88]} & 39 & \ci{[27, 49]} & 102.4 & \costbar{102.4} & & 81 & \ci{[75, 88]} & 27 & \ci{[21, 37]} & 99.2 & \costbar{99.2} \\
\arrayrulecolor{cHead}\specialrule{1pt}{0pt}{0pt}\arrayrulecolor{black}
\end{tabularx}
\begin{tablenotes}[flushleft]
\scriptsize
\item[] Bars show serial cost on a common linear scale; a full bar represents 102.4\,s.
\end{tablenotes}
\end{threeparttable}
\end{table*}

%% file: sections/discussion.tex
\section{Discussion and Conclusion}\label{sec:discussion}

\paragraph{Implications for reconstruction and data release}
On these data, speed and steering with a road map recover five-minute routes with errors of tens of metres.
Offline beam search reduces large failures over longer windows and needs the initial heading less than online
tracking does. For online use, the map filter and OPD-II have similar paired median errors at five minutes, while
the filter recovers more routes and has fewer large errors. With speed alone, Elastic Pathing benefits from
speed-based turn constraints that the benchmark filter and beam omit. These comparisons favour matching the
method to the available signals and information, rather than choosing a more complex motion model by default.

For data publishers, removing GPS coordinates or initial heading does not remove route information. The continuous
accelerator pedal adds little to speed and steering in our controlled comparison, but this does not establish that
pedal or speed releases are safe. Quantising steering to 10$^\circ$ or speed to 1\,m/s leaves the tested map attacks
largely effective. Evaluating a protection method also requires measuring its effect on the intended downstream
task and testing attackers adapted to the modified signals~\cite{raval2019olympus,buchholz2022raopt,yadav2024transprotect}.

\paragraph{Scope and limitations}
CANTraj broadens the range of drivers and vehicles available for CAN trajectory evaluation, but North American
roads and vehicles supported by the logging system dominate. Drivers are separate across splits; vehicle platforms
and regions are not. Road-density and turn-count comparisons are associations because road conditions, drivers and
vehicles vary together. Method rankings change on the controlled supplement (Section~\ref{sec:rq1}), so they should
not be treated as universal. Continuous pedal and CAN yaw signals are available on smaller subsets.

The task mainly measures recovery after one position is known. The area variants receive a square centred on that
position and are not tests of unrestricted unknown-start tracking; many routes leave the square. OPD-II's Meili
map access also differs from the disc used by the other map methods. Checks against the authors' samples support
the implementation of MacroTracking and OPD-II, but do not reproduce every original experiment. The other four
attacks lack the same validation evidence. Failure of these methods gives no privacy guarantee, and the
per-window oracle is a diagnostic result, not a deployable selector.

\paragraph{Reproducibility}
CANTraj and CANTraj-Bench contribute paired CAN--GNSS data, a common evaluation of trajectory recovery, and evidence
on the roles of steering, maps, prior information and decoding. The results show substantial location leakage from
vehicle signals and substantial variation across methods and driving conditions. Data, code and evaluation files
are available at \url{https://huggingface.co/datasets/HenryYHW/CANTraj} and
\url{https://github.com/WangYuHang-cmd/GPS--CAN}.